\documentclass[12pt]{article}
\usepackage{latexsym}
\usepackage{exscale}
\usepackage{graphicx}
\usepackage{amsmath}
\usepackage{amssymb}
\usepackage{natbib}
\renewcommand{\cite}{\citep}
\begin{document}
%
%
\title{\bf LAYER--RESOLVED MAGNETO--OPTICAL KERR EFFECT IN
SEMI--INFINITE INHOMOGENEOUS LAYERED SYSTEMS}
\author{A.~VERNES$^{\, a)}$
\footnote{Corresponding author. 
Tel.\ +43-1-58801-15840; fax: ++43-1-58801-15898.
\newline E-mail: av@cms.tuwien.ac.at},
L.~SZUNYOGH$^{\, a,b)}$ \\ and \\ P.~WEINBERGER$^{\, a)}$ \\
\ \\
$^{a)}$ {\small \it Center for Computational Materials Science}, \\
 {\small \it Technical University Vienna}, \\
 {\small \it Gumpendorferstr. 1a, 1060 Vienna, Austria} \\
$^{b)}$  {\small \it Department of Theoretical Physics }, \\
 {\small \it Budapest University of Technology and Economics} \\
 {\small \it Budafoki \'{u}t 8, 1521 Budapest, Hungary}}
\renewcommand{\today}{\textbf{submitted to Phase Transitions,\\
eds. P. Entel and W. Kleemann (Dec. 5, 2000)}}
\maketitle
%
%
\begin{abstract}
The contour integration technique applied to calculate the optical
conductivity tensor at finite temperatures in the case of
inhomogeneous surface layered systems within the framework of the
spin--polarized relativistic screened Korringa--Kohn--Rostoker band
structure method is extended to arbitrary polarizations of the
electric field. It is shown that besides the inter--band contribution,
the contour integration technique also accounts for the intra-band
contribution too. Introducing a layer--resolved complex Kerr angle,
the importance of the first, non--magnetic buffer layer below the
ferromagnetic surface on the magneto--optical Kerr effect in the
$\mathrm{Co} \mid \mathrm{Pt} _m$ multilayer system is shown.
Increasing the thickness of the buffer Pt, the layer--resolved complex
Kerr angles follow a linear dependence with respect to $m$ only after
nine Pt mono--layers.
\end{abstract}
%
\section{Introduction}
\label{sect:intro}
%
The magneto--optical Kerr effect (MOKE) was discovered by Rev. J.~Kerr
in 1876. Nowadays, the MOKE occurring in multilayers is investigated
because of obvious technological implications for high--density
magneto--optic recording media \cite{BS94,Man95}.

The first ab--initio calculation of the absorptive part of the optical
conductivity tensor based on the Kubo linear response theory
\cite{Kub57}, was carried out for Ni by \citet{WC74}. The dispersive
part of the optical conductivity tensor, needed to get the theoretical
MOKE, is usually computed using the Kramers--Kronig relations
\cite{DMAB88}. However, the first magneto--optical Kerr spectra,
namely those for Fe and Ni, were calculated on the basis of a
relativistic band--structure method by \citet{OMSK92}.

Commonly, the MOKE calculations for multilayers are performed using
conventional band--structure methods and super--cells
\cite{GE95,UUY+96}.  The contour integration technique developed by
two of the authors \cite{SW99} used in connection with the
spin--polarized relativistic screened Korringa--Kohn--Rostoker (SKKR)
method \cite{SUWK94,SUW95,USW95}, on the other hand, is a more
realistic approach for the magneto--optical Kerr spectra calculations
in case of surface layered systems.

In the next section (Section\ \ref{sect:theory}) the theoretical
framework is revisited and extended to arbitrary polarizations of the
electric field. In Section\ \ref{sect:continttech} it is demonstrated
that besides the inter--band contribution to the optical conductivity
tensor, the contour integration technique \cite{SW99} includes also
the intra--band contribution. In Section\ \ref{sect:limits} it is
shown that the Luttinger formalism \cite{Lut67} -- and accordingly the
contour integration technique too -- provides the complex conductivity
tensor even in the zero frequency limit for a finite life--time
broadening. The symmetric part of the dc conductivity tensor, however,
cannot be obtained by contour integrating in the complex energy plane.
Recently, the authors discussed in details schemes of how to control
the accuracy of the computation \cite{VSW00a}, therefore in Section\
\ref{sect:compdetails} only the main aspects of these numerical
methods are briefly reviewed. In Section\ \ref{sect:results} the
layer--resolved complex Kerr angles are introduced.  This concept is
then used to study separately the impact of the ferromagnetic surface
layer and of the non--magnetic buffer layers below the surface in the
$\mathrm{Co} \mid \mathrm{Pt} _m$ surface layered system. In addition,
the change of MOKE with the thickness of the buffer Pt--layer for a
particular frequency is also analyzed. Finally, the main results of
the present work are summarized in Section\ \ref{sect:summary}.

%
\section{Theoretical framework}
\label{sect:theory}
%

When a time dependent external electric field is applied to a solid,
currents are induced, which create internal electric fields. The total
electric field is then the sum of all these fields and in the
long--wavelength limit is written as \cite{Eyk86}
\begin{equation}
\Vec{E}(\Vec{r},t)=\Vec{E}_{0}\ e^{i(\Vec{q}\,\Vec{r}-\omega t)}\ 
e^{\delta t} \ , 
\label{eq:elpw}
\end{equation}
the positive infinitesimal $\delta $ implying the field to be turned
on at $t=-\infty $ \cite{Lax58}.

Because the solid is not a homogeneous system, in linear response the
resulting total current density is given by \cite{Mah90}
\begin{equation}
J_{\mu }(\Vec{r},t )= \sum_{\nu } \int d^3 r^{\prime} \int d
t^{\prime}\ \Sigma _{\mu \nu }(\Vec{r},\Vec{r}^{\,\prime};
t,t^{\prime}) E_{\nu }(\Vec{r}^{\,\prime},t^{\prime}) \ .
\label{eq:nonlocsgm}
\end{equation}
Here $\Sigma _{\mu \nu }(\Vec{r},\Vec{r}^{\,\prime}; t,t^{\prime})$ is
the non--local, time dependent conductivity and expresses the linear
response of the current at $t$ in $\Vec{r}$ and direction $\mu$ to the
local electric field applied at $t^{\prime}$ in $\Vec{r}^{\,\prime}$
and direction $\nu$ \cite{BZN+94}. 

The Fourier transform of Eq.\ (\ref{eq:nonlocsgm})
\begin{equation*}
J_{\mu }(\Vec{q}^{\,\prime },\omega ^{\prime })=\delta
_{\Vec{q}^{\,\prime },\Vec{q}}\delta (\omega ^{\prime }-\omega
)\sum_{\nu }\Sigma _{\mu \nu }( \Vec{q},\omega ) E_{0\nu } \qquad
(\mu,\nu=\mathrm{x,y,z})\ ,
\end{equation*}
introduces the wave--vector and frequency dependent complex
conductivity tensor $\boldsymbol{\Sigma}(\Vec{q} ,\omega )$
\cite{Eyk86}, which can be evaluated using the well--known Kubo
formula \cite{Kub57}
\begin{equation}
\Sigma _{\mu \nu }(\Vec{q},\omega )=\lim_{\delta \rightarrow
0^{+}}\frac{1}{V }\int_{0}^{\infty }dt\ e^{\frac{i}{\hbar }\left(
\hbar \omega +i\delta \right) t}\int_{0}^{\beta }d\lambda \
\left\langle J_{-\Vec{q}}^{\nu }\ J_{ \Vec{q}}^{\mu }\left( t+i\hbar
\lambda \right) \right\rangle _{\mathrm{eq}} \ .
\label{eq:kubo}
\end{equation}
Here $V$ is the crystalline volume, $\beta =\left( k_{\mathrm{B}
}T\right) ^{-1}$ and $\langle\ldots\rangle _{\mathrm{eq}}$ the
ensemble average in the equilibrium state at $t=-\infty$,
respectively.  Notice that when the electric field is not transverse,
i.e.\ $\Vec{q}\Vec{E}_0 \neq 0$ \cite{Eyk86,Mah90}, the Kubo conductivity
tensor $\boldsymbol{\Sigma}(\Vec{q},\omega )$ differs from that
derived from the Maxwell equations \cite{Kub66}, i.e.\
\begin{equation*}
\boldsymbol{\Sigma}_{\mathrm{M}}(\Vec{q},\omega ) =
\boldsymbol{\Sigma}(\Vec{q},\omega ) \left[ 1-\dfrac{4\pi i}{\omega} \
\hat{q}\cdot \boldsymbol{\Sigma}(\Vec{q},\omega ) \cdot \hat{q}
\right]^{-1} \ ,
\end{equation*}
where $\hat{q}$ is the unit vector along $\Vec{q}$.

\subsection{Luttinger formalism}
\label{sect:luttinger}

Based on the contour deformation method \cite{Hu93}, it can be easily
shown that
\begin{equation*}
\int_{0}^{\beta }d\lambda \ \left\langle J_{-\Vec{q}}^{\nu }\
J_{\Vec{q} }^{\mu }\left( t+i\hbar \lambda \right) \right\rangle
_{\mathrm{eq} }= \frac{i}{ \hbar }\int_{t}^{\infty }dt^{\prime }\
\left\langle \left[ J_{\Vec{q}}^{\mu }\left( t^{\prime }\right) ,\
J_{-\Vec{q}}^{\nu }\right] \right\rangle _{ \mathrm{eq}}\ ,
\end{equation*}
where $J_{\Vec{q}}^{\mu }\left( t^{\prime }\right)$ is the Heisenberg
operator:
\begin{equation*}
J_{\Vec{q}}^{\mu }\left( t^{\prime
}\right)  =e^{\frac{i}{\hbar }Ht^{\prime }}\ J_{\Vec{q}}^{\mu }\
e^{-\frac{i}{\hbar }Ht^{\prime }}
\ .  
\end{equation*}
For a diagonal representation of the one--electron Hamiltonian $H$ and
for the equilibrium density operator
\begin{equation*}
\rho _{\mathrm{eq}} \equiv f(H) =\dfrac{1}{ e^{\beta (H-\mu)}+1} \ ,
\end{equation*}
with $f(H)$ being the Fermi--Dirac distribution function, $\mu$ the
chemical potential (in the following the temperature dependence of the
chemical potential is neglected, i.e.\ $\mu$ equals the Fermi level
$\varepsilon _{ \mathrm{F}}$), Eq.\ (\ref{eq:kubo}) can be rewritten
as
\begin{equation}
\begin{aligned}
\Sigma _{\mu \nu }(\Vec{q},\omega )& = \lim_{\delta \rightarrow
0^{+}}\frac{i}{ \hbar V}\sum_{m,n}\ \left[ f(\varepsilon
_{m})-f(\varepsilon _{n})\right] \ J_{\Vec{q},mn}^{\mu }\
J_{-\Vec{q},nm}^{\nu } \\ & \cdot \int_{0}^{\infty }dt\
e^{\frac{i}{\hbar }\left( \hbar \omega +i\delta \right) t}
\int_{t}^{\infty }dt^{\prime }\ e^{\frac{i}{\hbar }(\varepsilon
_{m}-\varepsilon _{n})t^{\prime }}\ ,
\end{aligned}
\label{eq:sgminttt}
\end{equation}
where
\begin{equation*}
J_{\Vec{q},mn}^{\mu }\equiv \left\langle m\right| J_{\Vec{q}}^{\mu
}\left| n\right\rangle \ , \quad J_{-\Vec{q},nm}^{\nu }\equiv
\left\langle n\right| J_{-\Vec{q}}^{\nu }\left| m\right\rangle \ ,
\end{equation*}
and $\varepsilon _{m}$ denotes the eigenvalue of $H$ corresponding to
the eigenvector $\left| m\right\rangle$.

Both integrations in Eq.\ (\ref{eq:sgminttt}) with respect to
$t^{\prime } $ and $t$, respectively, can be performed taking
advantage on the Laplace transform of the identity
\cite{AS72}. Proceeding by this manner, the factor $e^{\delta t}$
introduced in Eq.\ (\ref{eq:elpw}) to describe the interaction of the
solid with its surroundings, can be seen also as a simple convergence
factor \cite{Lax58}. However, the complex conductivity tensor a for
finite $\zeta = \hbar \omega +i\delta$, i.e.\
\begin{equation}
\Sigma _{\mu \nu }(\Vec{q},\zeta ) =\frac{ \hbar }{iV}\sum_{m,n}\
\frac{f(\varepsilon _{m})-f(\varepsilon _{n})}{\varepsilon
_{m}-\varepsilon _{n}}\ \frac{J_{\Vec{ q},mn}^{\mu }\
J_{-\Vec{q},nm}^{\nu }}{\varepsilon _{m}-\varepsilon _{n}+\zeta } \ ,
\label{eq:sigmaqw}
\end{equation}
extends further the meaning of $\delta$. Here, $\delta$ can be seen as
the finite life--time broadening, which accounts for all the
scattering processes at finite temperature, which are not part of a
standard band--structure calculation. Introducing the current--current
correlation function as \cite{Mah90}
\begin{equation}
\sigma _{\mu \nu }\left( \Vec{q},\zeta \right) =\frac{i\hbar }{V}\sum
_{m,n}\ \frac{f(\varepsilon _{m})-f(\varepsilon _{n}) }{\varepsilon
_{m}-\varepsilon _{n}+\zeta }\ J_{\Vec{q},mn}^{\mu }\ J_{-\Vec{
q},nm}^{\nu } \quad , \label{eq:sigmazeta}
\end{equation}
finally results the well--known Luttinger formula \cite{Lut67}
\begin{equation}
\Sigma _{\mu \nu }(\Vec{q},\zeta) =\frac{\sigma _{\mu \nu }\left(
 \Vec{q} ,\zeta \right) -\sigma _{\mu \nu }\left( \Vec{q},0\right)
 }{\zeta } \quad . \label{eq:sgmzeta}
\end{equation}
A rather technical point to be noticed is that in a previous paper
\cite{SW99}, the variable $\zeta $ has a slightly different
definition, namely it was introduced as short--hand notation for
$\omega + i \delta / \hbar$. For this meaning of $\zeta$, one has to
use Eqs.\ (2) and (3) from \cite{SW99} instead of Eqs.\
(\ref{eq:sigmazeta}) and (\ref{eq:sgmzeta}).

The Luttinger formula (\ref{eq:sgmzeta}) differs in form from that
originally deduced by Kubo using the scalar potential description of
the electric field \cite{Kub57}. In fact, Eq.\ (\ref{eq:sigmazeta})
can be straightforwardly obtained starting from the vector potential
description of the electric field \cite{Cal74}. Nevertheless, the
formulae arising from the different descriptions of the electric field
are completely equivalent \cite{Hu93}.

The inter--band contribution to the complex conductivity tensor
$\Sigma _{\mu \nu }^{\mathrm{inter}}(\Vec{q},\zeta )$, can be obtained
directly from Eqs.\ (\ref{eq:sigmazeta}) and (\ref{eq:sgmzeta}),
imposing for the former $\varepsilon_m\neq \varepsilon_n$.
Considering the intra--band life--time broadening equal to that taken
for the inter--band contribution, Eq.\ (\ref{eq:sigmaqw}) provides in
the $\varepsilon _{m}\rightarrow \varepsilon _{n}$ limit immediately
the intra--band contribution
\begin{equation}
\Sigma _{\mu \nu }^{\mathrm{intra}}(\Vec{q},\zeta )= \dfrac{\sigma
_{\mu \nu }^{\mathrm{intra}}(\Vec{q})}{\zeta} \ ,
\label{eq:intrasgmzeta}
\end{equation}
where
\begin{equation}
\sigma _{\mu \nu }^{\mathrm{intra}}(\Vec{q}) \equiv -
\lim_{\varepsilon _{m}\rightarrow\varepsilon _{n}} \sigma _{\mu \nu
}(\Vec{q},0) \ , \label{eq:intrasigmaq}
\end{equation}
because $\sigma _{\mu \nu }^{\mathrm{intra}}\left( \Vec{q},\zeta
\right)= 0$, if $\zeta\neq 0$.  In contrast to the phenomenological
Drude term, which is supposed to give the intra--band contribution
\cite{OMSK92}, $\Sigma _{\mu \nu }^{\mathrm{intra}}(\Vec{q},\zeta )$
is added to both, diagonal ($\mu =\nu $) and off--diagonal ($\mu \neq
\nu $) elements of the complex conductivity tensor \cite{Opp99}.  In
conclusion, calculating $\sigma _{\mu \nu }\left( \Vec{q},\zeta
\right)$ in accordance with Eq.\ (\ref{eq:sigmazeta}), both the
inter-- and intra--band contribution is present in the Luttinger
formula (\ref{eq:sgmzeta}), i.e.\
\begin{equation*}
\Sigma _{\mu \nu }(\Vec{q},\zeta) = \Sigma _{\mu \nu
}^{\mathrm{inter}}(\Vec{q},\zeta) + \Sigma _{\mu \nu
}^{\mathrm{intra}}(\Vec{q},\zeta) \ .
\end{equation*}

\subsection{Contour integration technique}
\label{sect:continttech}

The contour integration technique, originally developed for
$\Vec{q}=0$ by \citet{SW99}, is extended below to calculate the
current--current correlation function for arbitrary polarizations of
the electric field. This technique permits the evaluation of Eq.\
(\ref{eq:sigmazeta}) by performing a contour integration in the
complex energy plane at finite temperature. The integration can be
performed by exploiting the properties of the Fermi--Dirac
distribution in the selection of the contour \cite{Mah90},
\begin{equation*}
f(z)=\frac{1}{e^{\beta(z-\mu)}+1} \quad \ ,
\end{equation*}
with $z\in \mathbb{C}$. Namely, that $f(z)$ is analytical everywhere,
except the so--called Matsubara poles \cite{NSW+94}
\begin{equation*}
z_{k} =\varepsilon_{\mathrm{F}}+i\left( 2k-1\right) \delta_{T} \ ,
\qquad k=0,\pm1,\pm2,\ldots
\end{equation*}
($\delta_{T} =\pi k_{\mathrm{B}}T$) and for energies parallel to the
real axis situated in--between two, successive Matsubara poles
\cite{WLZ+95}, $f(\varepsilon\pm2ik\delta_{T})=f(\varepsilon)$.

\begin{figure}[htbp] \centering
\includegraphics[width=0.7\columnwidth,clip]{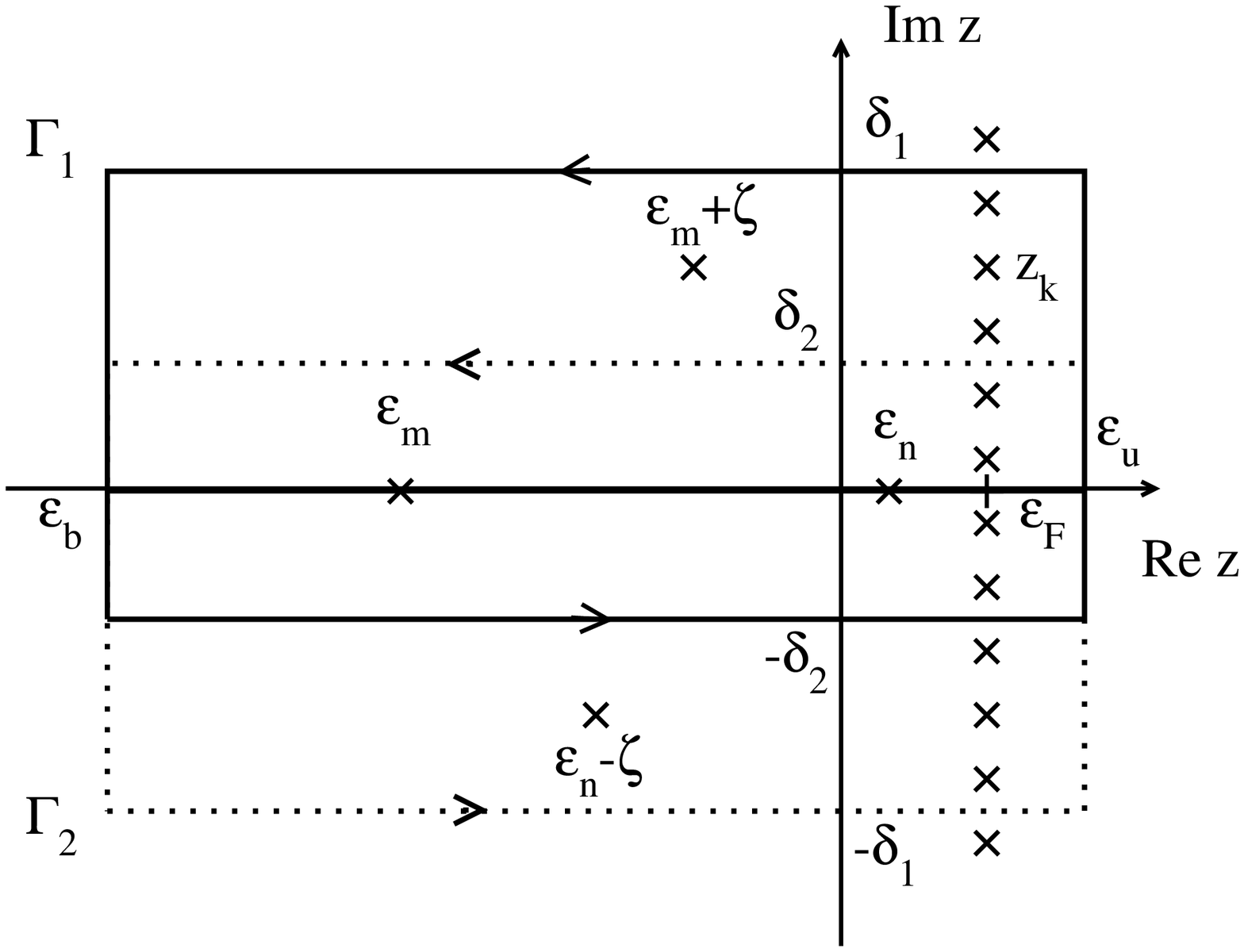} 
\caption[Selection of contours.]  
    {\label{fig:selofconts} 
    Selection of contour $\Gamma_1$ and $\Gamma_2$.  $\varepsilon_m$
    and $\varepsilon_n$ are eigenvalues of the Hamiltonian, $\zeta =
    \hbar\omega + i\delta$, with $\omega$ being the frequency and
    $\delta$ the life--time broadening.  $\varepsilon_{\mathrm{F}}$ is
    the Fermi level, $\varepsilon_{\mathrm{b}}$ the bottom valence
    band energy and $\varepsilon_{\mathrm{u}} =
    \varepsilon_{\mathrm{F}} + \xi k_{\mathrm{B}}T$
    ($\xi\in\mathbb{N}$). $z_k =\varepsilon_{\mathrm{F}}+i\left(
    2k-1\right) \delta_{T}$ are Matsubara poles with
    $k=0,\pm1,\pm2,\ldots$ and $\delta_{T} =\pi
    k_{\mathrm{B}}T$. $\delta_{1,2}$ were selected to fulfill the
    condition given by Eq.\ (\ref{eq:imagzpar}).}
\end{figure}

Initially, two contours $\Gamma_1$ and $\Gamma_2$ are considered
around the eigenvalues $\varepsilon_m$ and $\varepsilon_n$ including a
finite number of Matsubara poles, see Fig.\ \ref{fig:selofconts}. The
contour parts of $\Gamma_1$ parallel to the real axis
($z=\varepsilon\pm i\delta_{j}$, $j=1,2$) are taken in--between two,
successive Matsubara poles
\begin{equation}
(2N_{j}-1)\delta_{T} < \delta_{j} < (2N_{j}+1)\delta_{T} \ ,
\label{eq:imagzpar}
\end{equation}
with $N_{1}$ being the Matsubara poles in the upper and $N_{2}$ in the
lower semi--plane, respectively, included in $\Gamma_1$, e.g.\
$\delta_{j}=2N_{j}\delta_{T}$ for $j=1,2$.  $\Gamma_{2}$ is
$\Gamma_{1}$ mirrored along the real axis, hence includes $N_2$
Matsubara poles in the upper and $N_1$ in the lower semi--plane.  In
order to not have $\varepsilon_{n}-\zeta$ and $\varepsilon_{m} +\zeta$
contained in the contour $\Gamma_{2}$ and $\Gamma_{1}$, respectively,
the only constraint applied is
\begin{equation}
\delta>\delta_{2}\ .
\label{eq:constraint}
\end{equation}
%
%

Using the residue theorem, it was shown that \cite{SW99}
\begin{equation*}
\begin{aligned}
i\frac{f(\varepsilon_{m})}{\varepsilon_{m}-\varepsilon_{n}+\zeta} = &
-\frac{1}{2\pi}\int_{\Gamma_{1}}\hspace{-3.5ex}\circlearrowright\
dz\frac {f(z)}{\left( z-\varepsilon_{m})(z-\varepsilon_{n}+\zeta\right) } \\ 
& +i\frac{ \delta_{T}}{\pi}\sum_{k=-N_{2}+1}^{N_{1}}\frac{1}{(z_{k}-
\varepsilon_{m})(z_{k}-\varepsilon_{n}+\zeta)}
\end{aligned}
\end{equation*}
and
\begin{equation*}
\begin{aligned}
-i\frac{f(\varepsilon_{n})}{\varepsilon_{m}-\varepsilon_{n}+\zeta} & =
\frac {1}{2\pi}\int_{\Gamma_{2}}\hspace{-3.25ex}\circlearrowleft\
dz\frac {f(z)}{\left( z-\varepsilon_{n}\right) \left(
z-\varepsilon_{m}-\zeta\right) } \\
& +i\frac{\delta_{T}}{\pi}\sum_{k=-N_{1}+1}^{N_{2}}\frac
{1}{(z_{k}-\varepsilon_{n})(z_{k}-\varepsilon_{m}-\zeta)} \ .
\end{aligned}
\end{equation*}
Observing that as long as $\zeta \neq 0$, the sum of these two
expressions vanishes, when $\varepsilon_m=\varepsilon_n$, no
restriction regarding to the eigenvalues must be imposed, such that by
using the resolvent \cite{Wei90}
\begin{equation*}
G(z)=\sum_{n}\frac{\left| n\rangle\langle n\right|
}{z-\varepsilon_{n}}\ ,
\end{equation*}
immediately follows that
\begin{align}
\sigma_{\mu\nu}(\Vec{q},\zeta) & =\int_{\Gamma_{1}}\hspace
{-3.5ex}\circlearrowright\ dz\ f(z)\ \tilde{\sigma}_{\mu\nu}(
\Vec{q};z+\zeta,z)-\int_{\Gamma_{2}}\hspace{-3.25ex} \circlearrowleft\
dz\ f(z)\ \tilde{\sigma}_{\mu\nu}(\Vec {q};z,z-\zeta) \notag \\ &
-2i\delta_{T}\left[ \sum_{k=-N_{2}+1}^{N_{1}}\tilde{\sigma}_{\mu\nu }(
\Vec{q};z_{k}+\zeta,z_{k})+\sum_{k=-N_{1}+1}^{N_{2}}\tilde
{\sigma}_{\mu\nu}(\Vec{q};z_{k},z_{k}-\zeta)\right]
\label{eq:tildesgmqw} \ . 
\end{align}
Here due to the trace, the quantity
\begin{equation}
\tilde{\sigma}_{\mu\nu}(\Vec{q};z_{1},z_{2}) =-\frac{\hbar}{2\pi V}\
\mathrm{Tr}\left[ J_{\Vec{q}}^{\mu}\ G(z_{1})\ J_{-\Vec{q}}^{\nu}\
G(z_{2})\right] \ ,
\label{eq:tildesgmqz}
\end{equation}
contains the inter--band ($\varepsilon_m\neq \varepsilon_n$) and the
vanishing intra--band ($\varepsilon_m=\varepsilon_n$) contribution
too, i.e.\
\begin{equation}
\sigma_{\mu\nu}(\Vec{q},\zeta)=\sigma_{\mu\nu}^{\mathrm{inter}}(
\Vec{q},\zeta) \ .
\label{eq:sgmqzetasplitted}
\end{equation}
Thus the contour integration technique -- through Eq.\
(\ref{eq:tildesgmqw}) -- preserves all the features of the
current--current correlation function introduced by the Luttinger
formalism (Section\ \ref{sect:luttinger}).

Originally, $\tilde{\sigma}_{\mu\nu}(\Vec{q};z_{1},z_{2})$ was
introduced for $\Vec{q}=0$ as an auxiliary quantity to be evaluated
for the calculation of the residual resistivity ($\omega, T =0$) of
substitutionally disordered bulk systems \cite{But85}. Since its
extension to deal with inhomogeneously disordered layered systems
\cite{WLB+96}, this quantity nowadays is widely used to calculate the
magneto--transport properties of multilayers \cite{CWS+99}. As it was
shown in a previous paper \cite{SW99}, it also plays a central role,
when the magneto--optical properties of semi--infinite layered systems
are calculated.  In the case of electric fields with an arbitrary
polarization, $\tilde{\sigma}_{\mu\nu}(\Vec{q};z_{1},z_{2})$ obeys the
following symmetry relations
\begin{equation*}
\left\{ 
\begin{aligned}
\tilde{\sigma}_{\mu\nu}(\Vec{q};z_{1},z_{2}) & = \tilde{\sigma }
_{\nu\mu}(-\Vec{q};z_{2},z_{1}) \notag \\
\tilde{\sigma}_{\mu\nu}(\Vec{q};z_{1}^{\ast},z_{2}^{\ast}) & =
\tilde{\sigma}_{\nu\mu}^{\ast}(\Vec{q};z_{1},z_{2})=\tilde{\sigma
}_{\mu\nu}^{\ast}(-\Vec{q};z_{2},z_{1}) \ ,
\end{aligned}
\right.
\end{equation*}
by which Eq.\ (\ref{eq:tildesgmqw}) can be written as
\begin{align}
\sigma_{\mu\nu}(\Vec{q},\zeta) & =\int_{\Gamma_{1}}\hspace
{-3.5ex}\circlearrowright\ dz\ f(z)\ \tilde{\sigma}_{\mu\nu}(
\Vec{q};z+\zeta,z)-\left[ \int_{\Gamma_{1}}\hspace{-3.5ex}
\circlearrowright\ dz\ f(z)\ \tilde{\sigma}_{\mu\nu}(-\Vec
{q};z-\zeta^{\ast},z)\right] ^{\ast} \notag \\ &
-2i\delta_{T}\sum_{k=-N_{2}+1}^{N_{1}}\left[ \tilde{\sigma}_{\mu\nu }(
\Vec{q};z_{k}+\zeta,z_{k})+\tilde{\sigma}_{\mu\nu}^{\ast }(-
\Vec{q};z_{k}-\zeta^{\ast},z_{k})\right] \ .  \label{eq:sgmqw}
\end{align}
%

Although $\sigma_{\mu\nu}(\Vec{q},0)$ is obtained directly from Eq.\
(\ref{eq:sigmazeta}) by taking $\zeta = 0$, its evaluation requires
right from the beginning a single contour integration, namely that
along $\Gamma_1$, because the residue theorem provides directly
\begin{equation*}
\begin{aligned}
i\frac{f(\varepsilon_{m})-f(\varepsilon_{n})}{\varepsilon_{m}-\varepsilon_{n}
} = & -\frac{1}{2\pi}\int_{\Gamma_{1}}\hspace{-3.5ex}\circlearrowright \
dz\frac{ f(z)}{\left( z-\varepsilon_{m})(z-\varepsilon_{n}\right) } \\
& +i\frac{\delta_{T}
}{\pi}\sum_{k=-N_{2}+1}^{N_{1}}\frac{1}{(z_{k}-\varepsilon_{m})(z_{k}-
\varepsilon_{n})} \ .
\end{aligned}
\end{equation*}
>From this expression results immediately the inter-band contribution
$\sigma^{\mathrm{inter}}_{\mu \nu }(\Vec{q},0)$, when $\varepsilon_m
\neq \varepsilon_n$ and in the $\varepsilon_m\rightarrow\varepsilon_n$
limit this leads to
\begin{equation*}
i\frac{\partial f(\varepsilon _{n})}{\partial \varepsilon
_{n}}=-\frac{1}{ 2\pi }\int_{\Gamma
_{1}}\hspace{-3.5ex}\circlearrowright \ dz\frac{f(z)}{ (z-\varepsilon
_{n})^{2}}\ +i\frac{\delta _{T}}{\pi }
\sum_{k=-N_{2}+1}^{N_{1}}\frac{1}{(z_{k}-\varepsilon _{n})^{2}}\ .
\end{equation*}
Therefore, according to Eq.\ (\ref{eq:tildesgmqz}),
\begin{equation}
\sigma_{\mu\nu}(\Vec{q},0)=\int_{\Gamma_{1}}\hspace{-3.5ex}
\circlearrowright\ dz\ f(z)\ \tilde{\sigma}_{\mu\nu}(\Vec
{q};z,z)-2i\delta_{T}\sum_{k=-N_{2}+1}^{N_{1}}\tilde{\sigma}_{\mu\nu
}( \Vec{q};z_{k},z_{k})\  \label{eq:sgmq0}
\end{equation}
includes the intra--band contribution too, i.e.,\ based on Eq.\
(\ref{eq:intrasigmaq}),
\begin{equation}
\sigma _{\mu \nu }(\Vec{q},0)=\sigma _{\mu \nu }^{\mathrm{inter}}(%
\Vec{q},0)-\sigma _{\mu \nu }^{\mathrm{intra}}(\Vec{q})\ .
\label{eq:sgmq0splitted}
\end{equation}

In conclusion, it can be stated that within the contour integration
technique, the optical conductivity tensor obtained by using the
Luttinger formula (\ref{eq:sgmzeta}), contains both inter-- and
inter--band contributions, i.e.\ based on Eqs.\
(\ref{eq:sgmqzetasplitted}), (\ref{eq:sgmq0splitted}) and
(\ref{eq:intrasgmzeta}),
\begin{equation*}
\Sigma_{\mu\nu}(\Vec{q},\zeta )=
\dfrac{\sigma^{\mathrm{inter}}_{\mu \nu }(\Vec{q},\zeta )
-\sigma^{\mathrm{inter}}_{\mu \nu }(\Vec{q},0)}{\zeta }
+\Sigma^{\mathrm{intra}}_{\mu \nu }(\Vec{q}) \ .
\end{equation*}
%

\subsection{Static and sharp bands limit}
\label{sect:limits}

In order to derive the dc electrical conductivity, it has been shown,
that first the limit of $\Vec{q}\rightarrow0$ has to be taken
\cite{Lut64}.  Eqs.\ (\ref{eq:sigmazeta}) and (\ref{eq:sgmzeta}) lead
to finite values for $\Vec{q}\rightarrow0$ in the static regime
($\omega\rightarrow0$), or, separately, on sharp bands limit
($\delta\rightarrow0^{+}$). It should be recalled that the contour
integration technique can be only used for $\omega = 0$ calculations,
when
\begin{equation*}
\Sigma_{\mu\nu}\left(\delta\right)
=\left. \frac{\partial\sigma_{\mu\nu}(\zeta)}
{\partial\omega}\right|_{\omega=0}\ ,
\end{equation*}
because of the constraint (\ref{eq:constraint}) applied to the contour
$\Gamma_1$, the sharp bands limit ($\delta=0$) is unreachable even
when $T=0$ K.

Alternatively, in the static limit and for finite life--time
broadening, Eq.\ (\ref{eq:sgmqw}) is given by
\begin{equation*}
\Sigma_{\mu\nu}\left( \delta\right) =\frac{1}{2}\left[
\Sigma_{\mu\nu}^{(+)}\left( \delta\right) +\Sigma_{\mu\nu}
^{(-)}\left( \delta\right) \right] -\frac{1}{2}\left[
\Sigma_{\mu\nu}^{(+)}\left( \delta\right) -\Sigma_{\mu\nu}^{(-)}\left(
\delta\right) \right] \ ,
\end{equation*}
where
\begin{equation*}
\Sigma_{\mu\nu}^{(\pm)}\left( \delta\right) \equiv\mp\frac{\hbar}
{iV}\sum_{m,n}\frac{f(\varepsilon_{m})-f(\varepsilon_{n})}{\varepsilon
_{m}-\varepsilon_{n}}\frac{J_{mn}^{\mu}\ J_{nm}^{\nu} }{\left(
\varepsilon_{m}-\varepsilon_{n}\right) \mp i\delta} \ .
\end{equation*}
The sharp bands limit of its symmetric part \cite{Lut67}
\begin{equation*}
\Sigma_{\mu\nu}^{(\mathrm{s})} =\frac{1}{2}\left[
\Sigma_{\mu\nu}^{(+)} +\Sigma_{\mu\nu}^{(-)} \right] \ ,\qquad
\mbox{with}\quad \Sigma_{\mu\nu}^{(\pm)} =\lim_{\delta\rightarrow0^{+}
}\Sigma_{\mu\nu}^{(\pm)}\left( \delta\right) \ ,
\end{equation*}
on the other hand, exists, but is restricted to the intra--band
contribution and is given by the Landau form of the dc conductivity
\cite{Gre58}
\begin{equation*}
\Sigma_{\mu\nu}^{(\mathrm{s})} =\frac{\pi\hbar}{V} \sum_{m,n}\left[
-\frac{\partial f(\varepsilon_{n})}{\partial\varepsilon_{n} }\right] \
\delta\left( \varepsilon_{m}-\varepsilon_{n}\right) \ J_{mn}^{\mu}
J_{nm}^{\nu}\ .
\end{equation*}
As a consequence of the presence of the Dirac $\delta$--function in
this expression, $\Sigma_{\mu\nu}^{(\mathrm{s})}$ can be calculated
performing an energy integration along the real axis when $T > 0$,
i.e.,\
\begin{equation*}
\Sigma_{\mu\nu}^{(\mathrm{s})} =\int_{-\infty}^{\infty }d\varepsilon\
\left[ -\frac{\partial f(\varepsilon)}{\partial\varepsilon }\right] \
\sigma_{\mu\nu}\left( \varepsilon\right) \ ,
\end{equation*}
where $\sigma_{\mu\nu}\left( \varepsilon\right)$ is the arithmetic
mean of the four $\tilde{\sigma}_{\mu\nu}\left( z_1,z_2\right) $,
$z_{1,2}=\varepsilon\pm i0$, given by Eq.\ (\ref{eq:tildesgmqz}) in
the $\Vec{q}\rightarrow 0$ limit \cite{But85,WLB+96}.

\section{Computational details}
\label{sect:compdetails}
%

In principle, the contour $\Gamma_{1}$ along which one integrates in
Eqs.\ (\ref{eq:sgmqw}) and (\ref{eq:sgmq0}), should extend from
$+\infty$ to $-\infty$. In practice, however, the lower limit of the
contour $\Gamma_{1}$ is set to the bottom valence band energy
$\varepsilon _{\mathrm{b}}$, i.e.,\ no core states are assumed to
contribute. For the upper limit, on the other hand,
$\varepsilon_{\mathrm{u}} = \varepsilon_{\mathrm{F}} + \xi
k_{\mathrm{B}}T$ ($\xi\in\mathbb{N}$) instead of $\infty$ is taken,
because the Fermi--Dirac distribution decays fast. (For more details,
see Fig.\ \ref{fig:selofconts}.)

In calculating the difference
between Eq.\ (\ref{eq:sgmqw}) and (\ref{eq:sgmq0}), one can
distinguish between four different contributions, i.e.
\begin{equation*}
\sigma_{\mu\nu}(\Vec{q},\zeta) -  \sigma_{\mu\nu}(\Vec{q},0)
=\sum_{j=1}^{4}\sigma_{\mu\nu}^{(j)}( \Vec{q},\zeta)\ .
\end{equation*}
The part of contour $\Gamma_1$ in the upper semi--plane contributes
\begin{equation*}
\begin{aligned}
\sigma_{\mu\nu}^{(1)}(\Vec{q},\zeta) &
=\int_{\varepsilon_{\mathrm{b}}+i0}
^{\varepsilon_{\mathrm{u}}+i\delta_{1}}\ dz\ f(z)\ \left[
\tilde{\sigma}_{\mu\nu}(\Vec{q};z+\zeta,z)
-\tilde{\sigma}_{\mu\nu}(\Vec{q};z,z) \right] \\ & -\left[
\int_{\varepsilon_{\mathrm{b}}+i0}
^{\varepsilon_{\mathrm{u}}+i\delta_{1}}\ dz\ f(z)\
\tilde{\sigma}_{\mu\nu}(-\Vec{q};z-\zeta^{\ast},z)\right] ^{\ast}\ ,
\end{aligned}
\end{equation*}
whereas the contour part in the lower semi--plane contributes
\begin{equation*}
\begin{aligned}
\sigma_{\mu\nu}^{(2)}(\Vec{q},\zeta)
= & -\int_{\varepsilon_{\mathrm{b}}-i0}
^{\varepsilon_{\mathrm{u}}-i\delta_{2}}\ dz\ f(z)\ \left[
\tilde{\sigma}_{\mu\nu}(\Vec{q};z+\zeta,z) -\tilde{\sigma}
_{\mu\nu}(\Vec{q};z,z) \right] \\ & +\left[
\int_{\varepsilon_{\mathrm{b}}-i0}
^{\varepsilon_{\mathrm{u}}-i\delta_{2}}dz\ f(z)\
\tilde{\sigma}_{\mu\nu}(-\Vec{q};z-\zeta^{\ast},z)\right] ^{\ast}\ .
\end{aligned}
\end{equation*}
It should be mentioned, that in a previous paper the sign of the
latter was misprinted, see Eq.\ (25) from \cite{SW99}.  The Matsubara
poles have two contributions: one coming from the $N_{1}-N_{2}$ poles
situated in the upper semi--plane
\begin{equation*}
\begin{aligned}
\sigma_{\mu\nu}^{(3)}(\Vec{q},\zeta)=-2i\delta_{T}
\sum_{k=N_{2}+1}^{N_{1}} & \left[ \hspace{2ex} 
\tilde{\sigma}_{\mu\nu}(\Vec{q};z_{k}+\zeta,z_{k})
-\tilde{\sigma}_{\mu\nu}(\Vec{q};z_{k},z_{k}) \right. \\
& \left. 
+\tilde{\sigma}_{\mu\nu}^{\ast}(-\Vec{q};z_{k}-\zeta^{\ast},z_{k})
\hspace{2ex} \right] 
\end{aligned}
\end{equation*}
and an other one from the $2N_{2}$ poles near and on both sides of the
real axis
\begin{align*}
\sigma_{\mu\nu}^{(4)}(\Vec{q},\zeta)  = -2i\delta_{T} &
\sum_{k=1}^{N_{2}} \left[ 
\tilde{\sigma}_{\mu\nu}(\Vec{q};z_{k}+\zeta,z_{k})
-\tilde{\sigma}_{\mu\nu}(\Vec{q};z_{k},z_{k})
+\tilde{\sigma}_{\mu\nu}^{\ast}(-\Vec{q};z_{k}-\zeta^{\ast},z_{k})
\right.  \notag \\ & \left. 
+\tilde{\sigma}_{\mu\nu}(\Vec{q};z_{k}^{\ast}+\zeta,z_{k}^{\ast})
-\tilde{\sigma }_{\mu\nu}(\Vec{q};z_{k}^{\ast},z_{k}^{\ast})
+\tilde{\sigma}_{\mu\nu}^{\ast}(-\Vec{q};z_{k}^{\ast}-\zeta^{\ast},z_{k}^{\ast})
\right] \ .
\end{align*}

In the present paper, the current--current correlation function needed
to evaluate $\sigma_{\mu\nu}^{(j)}( \Vec{q},\zeta)$ is computed using
Eq.\ (\ref{eq:tildesgmqz}), relativistic current operators
\cite{WLB+96} and the Green functions $G(z)$ obtained by means of the
spin--polarized relativistic screened Korringa--Kohn--Rostoker (SKKR)
method for layered systems \cite{SUWK94,SUW95,USW95}. Because of the
finite imaginary part of the complex energy variable $z$, the
calculation scheme includes the so--called irregular solutions of the
Dirac equation too \cite{SW99}.

>From the computational point of view, besides the Matsubara poles, the
optical conductivity tensor $\Sigma_{\mu\nu}(\Vec{q},\zeta)$ as given
by Eq.\ (\ref{eq:sgmzeta}), depends also on the number of complex
energy points $n_z$ considered for the energy integrals involved
($\sigma_{\mu\nu}^{(1,2)}( \Vec{q},\zeta)$), on the number of
$\Vec{k}$--points used to calculate the scattering path operator and
the $\tilde{\sigma}_{\mu\nu}(\Vec{q};z \pm \hbar \omega + i \delta,z)$
for a given energy $z$, respectively.  Recently, the authors have
proposed two schemes to control the accuracy of the $z$-- and
$\Vec{k}$--integrations computing $\Sigma _{\mu\nu} (\Vec{q},\zeta )$
\cite{VSW00a}. For that reason, only the main aspects of these
numerical methods are given below.

The accuracy of the integrations with respect to $z$ is controlled
comparing the obtained results by means of the Konrod quadrature
\cite{Lau97,CGG+00}, $\mathcal{K}_{2n_z+1} \sigma_{\mu\nu}^{(j)}(
\Vec{q},\zeta) $, with those computed by using the Gauss integration
rule \cite{PFT+92}, $\mathcal{G}_{n_z} \sigma_{\mu\nu}^{(j)}(
\Vec{q},\zeta) $, on each contour part in both semi--planes ($j=1,2$)
-- for the notations used, see \citet{VSW00a}. Hence along a
particular contour part, $\sigma _{\mu \nu}^{(j)}( \Vec{q}, \zeta )$
is said to be converged, if
\begin{equation}
\max \ \left| \ \mathcal{K}_{2n_z+1}
\sigma_{\mu\nu}^{(j)}(\Vec{q},\zeta) - \mathcal{G}_{n_z}
\sigma_{\mu\nu}^{(j)}(\Vec{q},\zeta) \ \right| \; \leq \; \epsilon_z
\, ,
\label{eq:zconv}
\end{equation}
where $\epsilon_z$ is the accuracy parameter. One advantage of this
scheme is that the integrands have to be evaluated only in $2n_z+1$
points, because the $2n_z+1$ Konrod--nodes include all the $n_z$
Gauss--nodes.  Another advantage is, that Eq.\ (\ref{eq:zconv}) is
fulfilled for any, arbitrary small $\epsilon_z$, as test calculations
performed for $\Vec{q}=0$ have shown \cite{VSW00a}.

In order to compute the involved two--dimensional $\Vec{k}$--space
integrals with arbitrary high precision, a new, cumulative special
points method was developed by the present authors \cite{VSW00a}. This
method exploits the arbitrariness of the mesh origin \cite{Haw92} and
requires to evaluate the integrands only for the $\Vec{k}$--points
newly added to the mesh. Test calculations for $\Vec{q}=0$ have shown
\cite{VSW00a}, that a requirement similar to Eq.\ (\ref{eq:zconv}),
i.e.\
\begin{equation}
\max \ \left| \ \mathcal{S}_{n_i}
\tilde{\sigma}_{\mu\nu}(\Vec{q};z^{\prime},z)-
\mathcal{S}_{n_{i-1}}
\tilde{\sigma}_{\mu\nu}(\Vec{q};z^{\prime},z) \ \right| \; \leq \;
\epsilon_{\Vec{k}} \, ,
\label{eq:kconv}
\end{equation}
for any $z$ on the contour or $z_k$ Matsubara pole
($z^{\prime}=z+\zeta, z-\zeta^{\ast}$) can be fulfilled with arbitrary
high accuracy $\epsilon_{\Vec{k}}$.

Furthermore, it was shown \cite{VSW00a}, that if the $z$-- and
$\Vec{k}$--integrations are performed and controlled in the manner
presented above, the computed optical conductivity
$\Sigma_{\mu\nu}(\Vec{q},\zeta)$ does not depend on the form of the
contour in the upper semi--plane.  Hence our computational set--up for
$\Sigma_{\mu\nu}(\Vec{q},\zeta)$ is completely specified by
$\epsilon_z$, $\epsilon_{\Vec{k}}$ and the number of Matsubara poles
$N_2$ near and on both sides of the real axis. (The latter is taken in
accordance with the life--time broadening $\delta$, i.e.,\ to fulfill
the condition: $2\delta_2 = \delta$.)

%
\section{Results and discussions}
\label{sect:results}
%
The layered system studied in the present paper consists on a
mono--layer of Co on the top of fcc--Pt(100) with Pt--layers serving
as ``buffer'' to bulk Pt \cite{PZU+99}:
\begin{center}
Co$\, \mid \,$
Pt$_m \, \equiv \,$
Co$_{(1)}\, \mid \, $ 
Pt$_{(2)}\, \mid \, $ 
$\ldots\,\mid\,$
Pt$_{(n-1)}\,\mid\,$
Pt$_{(n)}\,\mid\,$
Pt (bulk)
\end{center}
with the subscript in parenthesis being the layer index. The bottom
valence band energy $\epsilon _{\mathrm{b}}$ was taken at $-1$ Ryd,
$\varepsilon_{\mathrm{u}} = \varepsilon_{\mathrm{F}} +
8k_{\mathrm{B}}T$ and the Fermi level $\epsilon _{\mathrm{F}}$ is that
of Pt bulk, i.e.\ $-0.039$ Ryd.  (Pt bulk acts as a charge reservoir
for the layered system.)

The optical conductivity calculations were carried out for
$\Vec{q}=0$, $T=300$ K and using a life--time broadening of $0.048$
Ryd, i.e.\ $N_2$ = 2 Matsubara poles near and on both sides of the
real axis.  The computation is less influenced by the Matsubara poles
considered in the upper semi--plane, as it was already mentioned in
Section\ \ref{sect:compdetails}, therefore we have taken $N_1-N_2=35$
poles to accelerate the computation in the upper semi--plane. The
convergence criteria (\ref{eq:zconv}) and (\ref{eq:kconv}) were
fulfilled for $\epsilon _z = \epsilon _{\Vec{k}} = 10^{-3}$ a.u. 

It was shown \cite{WLB+96}, that in case of layered systems the
$\tilde{\sigma} _{\mu\nu} (\Vec{q} ; z_{1}, z_{2} )$ over which one
has to integrate, see Section\ \ref{sect:continttech}, can be split
into intra-- ($p=q$) and inter--layer ($p \neq q$) contributions. This
decomposition of the current--current correlation function in Eq.\
(\ref{eq:tildesgmqz}), makes the optical conductivity for layered
systems to be of the form
\begin{equation}
\Sigma_{\mu\nu}(\Vec{q},\zeta) = \sum_{p=1}^{n}
\Sigma^{\,p}_{\mu\nu}(\Vec{q},\zeta)\ ,
\label{eq:Sgm}
\end{equation}
where the layer--resolved optical conductivities are given by
\begin{equation}
\Sigma^{\,p}_{\mu\nu}(\Vec{q},\zeta) = \sum_{q=1}^{n}
\Sigma^{pq}_{\mu\nu}(\Vec{q},\zeta)\ , \qquad p=1,\ldots,n \ . 
\label{eq:lrSgm}
\end{equation}
Because in the present work, $\Vec{q}=0$, $\delta$ and $T$ are fixed,
in the following these variables are omitted and the optical
conductivity tensor is simply denoted by $\Sigma_{\mu\nu}(\omega)$.
In the case of the layered system $\mathrm{Co}\mid\mathrm{Pt}_m$, for the
inter--layer contribution it is verified with an accuracy of
$10^{-15}$ $\mathrm{fs}^{-1}$ that
\begin{equation*}
\Sigma^{pq}_{\mu\nu}(\omega) = \Sigma^{qp}_{\mu\nu}(\omega)\ .
\end{equation*}
It was also found that these inter--layer contributions are always
smaller than the intra--layer contributions
$\Sigma^{pp}_{\mu\nu}(\omega)$.

When linearly polarized light is reflected from a magnetic solid, the
reflected light becomes elliptically polarized and its polarization
plane is rotated with an angle $\theta _{\mathrm{K}}$ with respect to
the incident light.  The former effect is characterized by the
ellipticity $\varepsilon _{\mathrm{K}}$ and the phenomenon is called
magneto--optical Kerr effect (MOKE). In the polar geometry, considered
in the present work, both the incident light and the magnetization are
perpendicular to the surface.  For a precision up to several degrees
the complex Kerr angle is given then by \cite{RS90}
\begin{equation}
\Phi _{\mathrm{K}} = \theta _{\mathrm{K}} + i \varepsilon
_{\mathrm{K}} = \dfrac{\Sigma _{\mathrm{xy}}(\omega)} {\Sigma
_{\mathrm{xx}}(\omega)} \left[ 1 + \dfrac{4\pi i}{\omega}\,\Sigma
_{\mathrm{xx}}(\omega) \right]^{-\frac{1}{2}} \ .
\label{eq:moke}
\end{equation}

\begin{figure}[hbtp] \centering
\begin{tabular}{cc}
\includegraphics[width=0.47\columnwidth,clip]{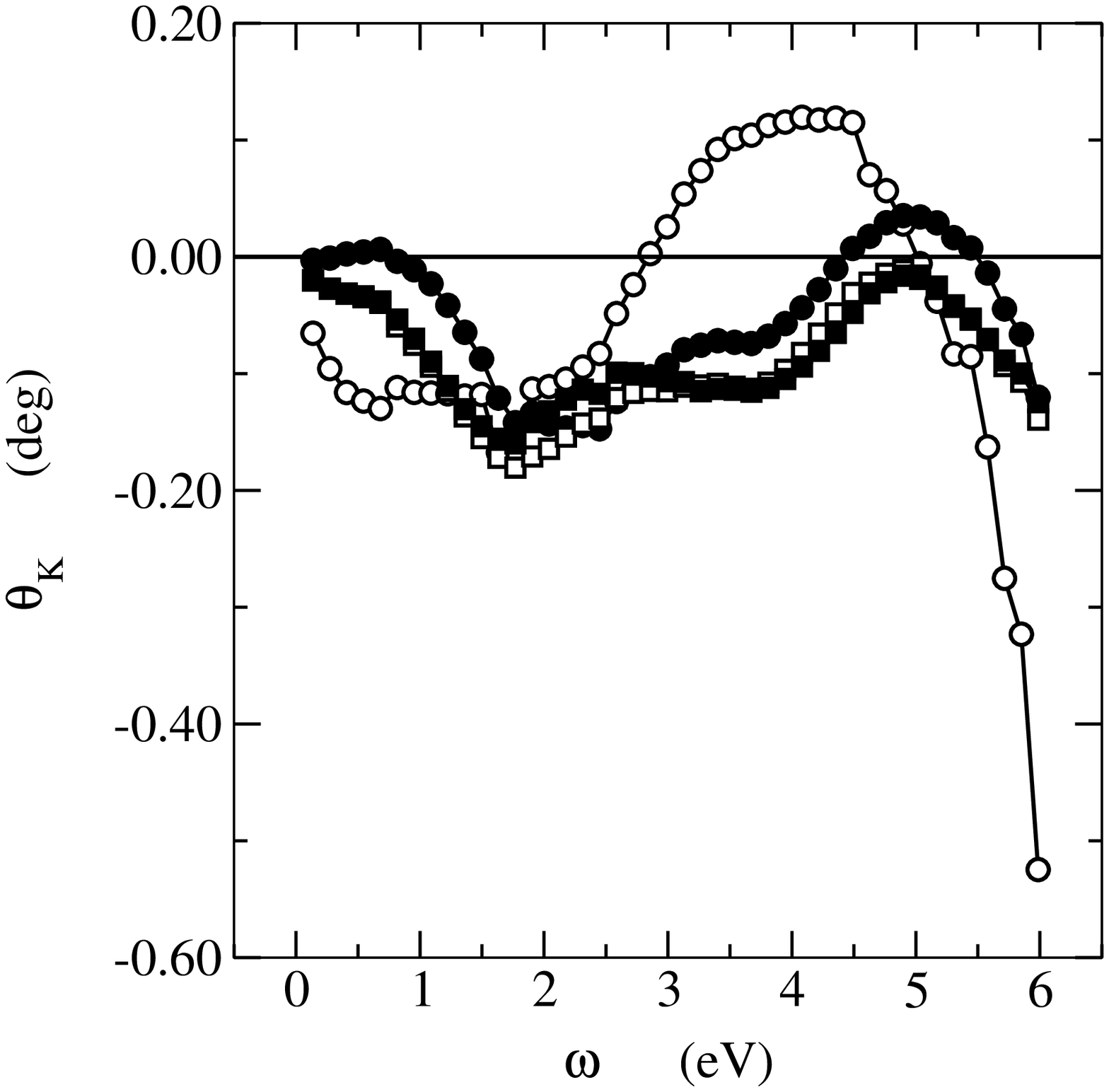} &
\includegraphics[width=0.47\columnwidth,clip]{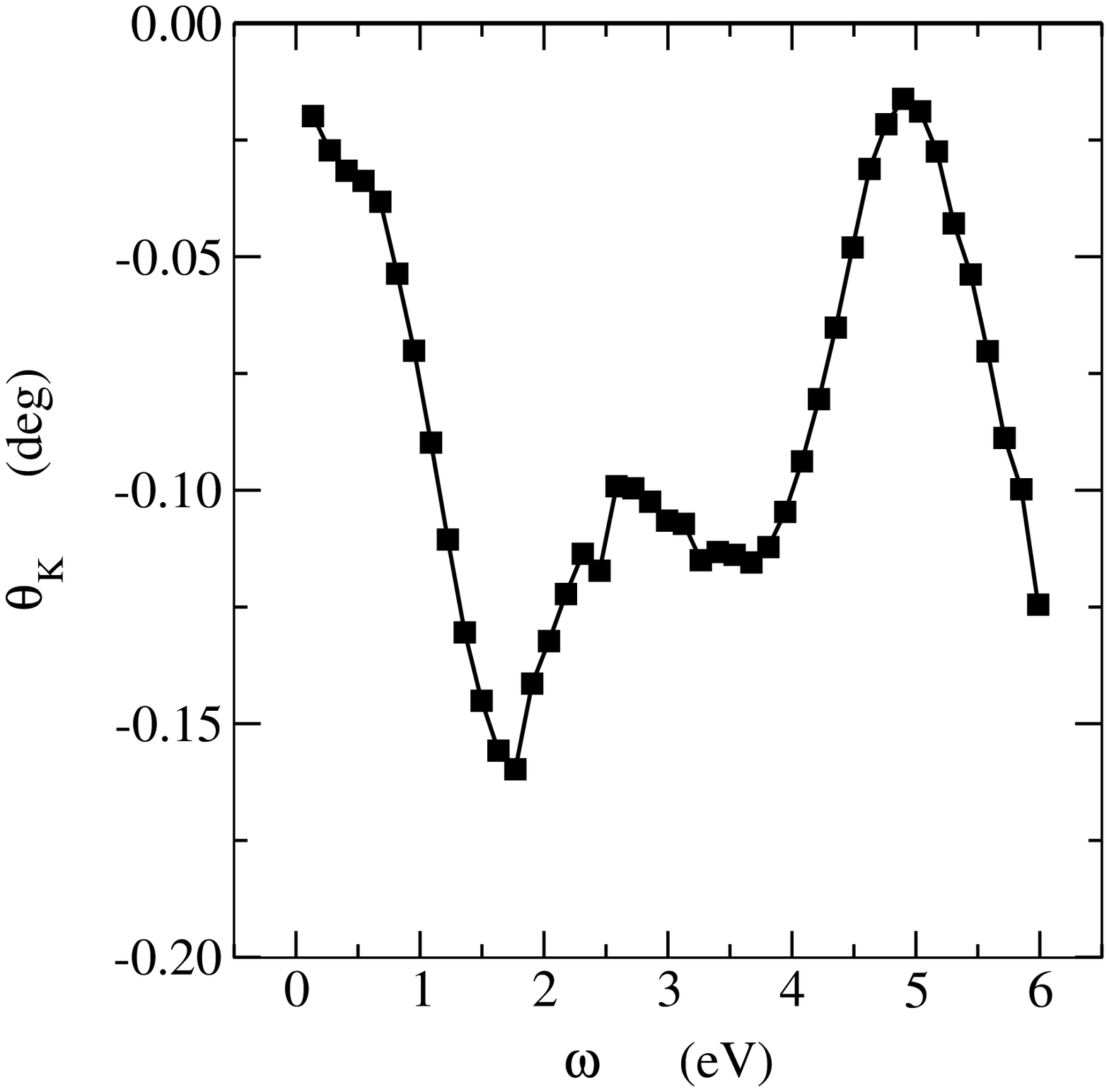} \\
\includegraphics[width=0.47\columnwidth,clip]{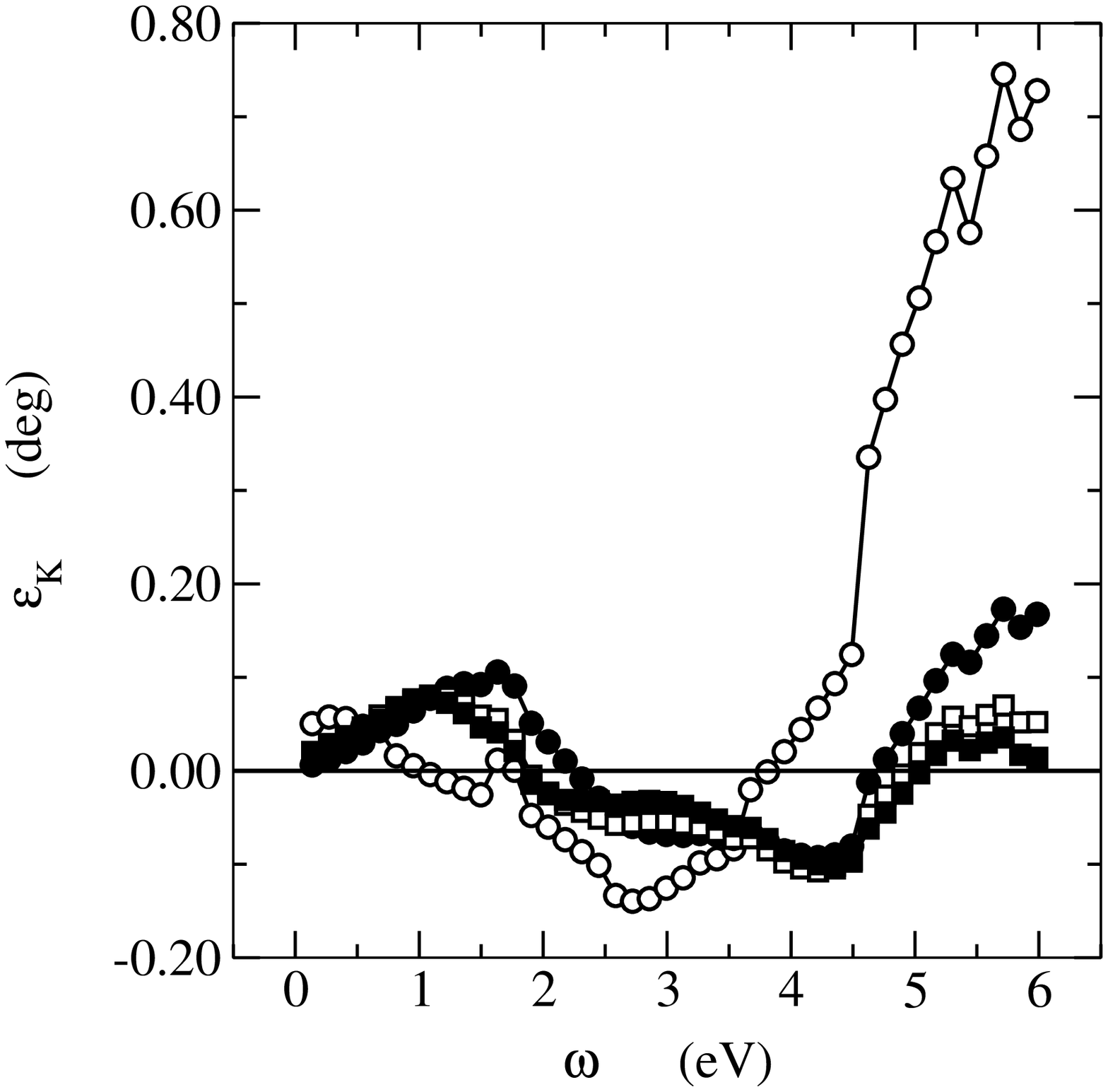} &
\includegraphics[width=0.47\columnwidth,clip]{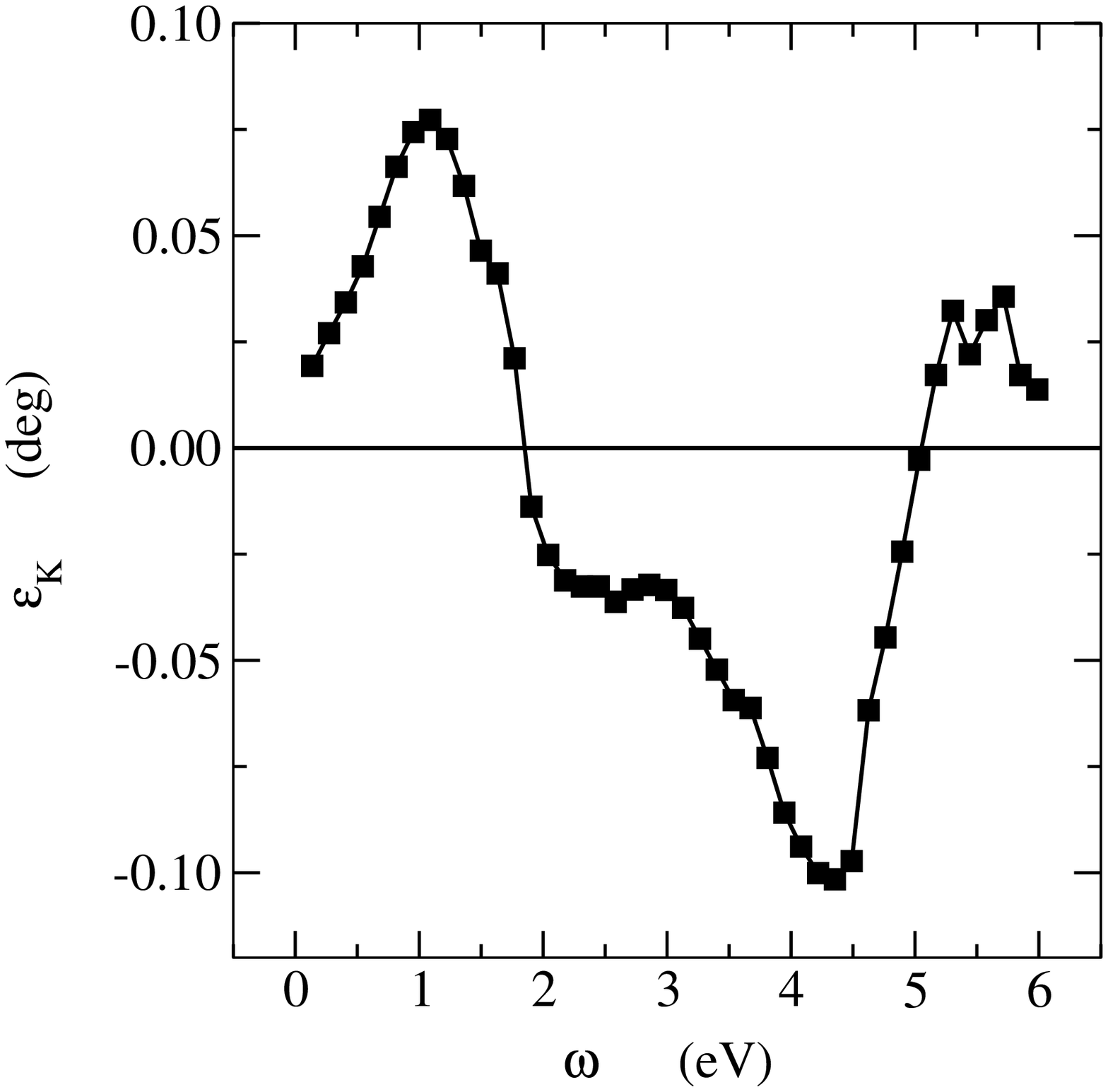} \\
\end{tabular}
\caption[$\protect n$--dependence of complex Kerr effect.]
    {\label{fig:ndepkerr}
      Complex Kerr effect for the $\mathrm{Co}\mid\mathrm{Pt}_m$
      layered system with $m=0\ (\circ)$, 1 ($\bullet$), 2
      ($\scriptstyle\square$), 3 ($\scriptstyle\blacksquare$)
      Pt--layers below the surface Co--layer (left panel) and
      separately for $m=3$ (right panel).  $\delta=0.048$ Ryd, $T$ =
      300 K and $\varepsilon_z = \varepsilon_{\vec{k}} = 10^{-3}$
      a.u., respectively.}
\end{figure}

The frequency dependence of $\theta _{\mathrm{K}}$ and $\varepsilon
_{\mathrm{K}}$ in case of $\mathrm{Co}\mid\mathrm{Pt}_m$ surface
layered system for different number of Pt buffer layers
($m=0,\ldots,3$) is shown in Fig.\ \ref{fig:ndepkerr}. As a general
trend of these curves, it can be observed, that apart from the $m=0$
situation at low photon energies, our calculations are in agreement
with the experimental fact, that the MOKE decreases, if the thickness
of the buffer Pt increases \cite{GVT+98}.

In fact, $m=0$ means, that one considers only the surface Co--layer in
the calculations, i.e.,\ Eqs.\ (\ref{eq:Sgm}) and (\ref{eq:lrSgm}) are
taken for $n=1$. Although in this case, the layer--resolved optical
conductivity equals $\Sigma _{\mu\nu} (\omega )$, the corresponding
Kerr spectrum is the most peculiar one in comparison with the results
obtained for $m\ge 1$. Once the contribution of different Pt buffer
layers below the surface are accounted for ($m\ge 1$), the complex
Kerr angle changes dramatically in the whole frequency range and the
calculated MOKE for $\mathrm{Co}\mid\mathrm{Pt}_{m=3}$, see right
panels in Fig.\ \ref{fig:ndepkerr}, possesses already the features
known from the experiments: $\theta _{\mathrm{K}}$ has two typical,
local minima about 2 and 4 eV and $\varepsilon _{\mathrm{K}}$ has a
flat region in--between two, local extrema in the middle of the
frequency range \cite{UUY+96}. The frequency width, however, is
smaller than that known from experiments. Below 1 eV and above 5 eV
our theoretical spectra are richer in fine details than the
experimental data \cite{EPY+98}. Finally, it should be mentioned that,
both the Kerr rotation angle and ellipticity are approximatively three
times smaller than those measured for thin films of Co--Pt alloy
\cite{Wel96}.

\begin{figure}[hbtp] \centering
\begin{tabular}{cc}
\includegraphics[width=0.47\columnwidth,clip]{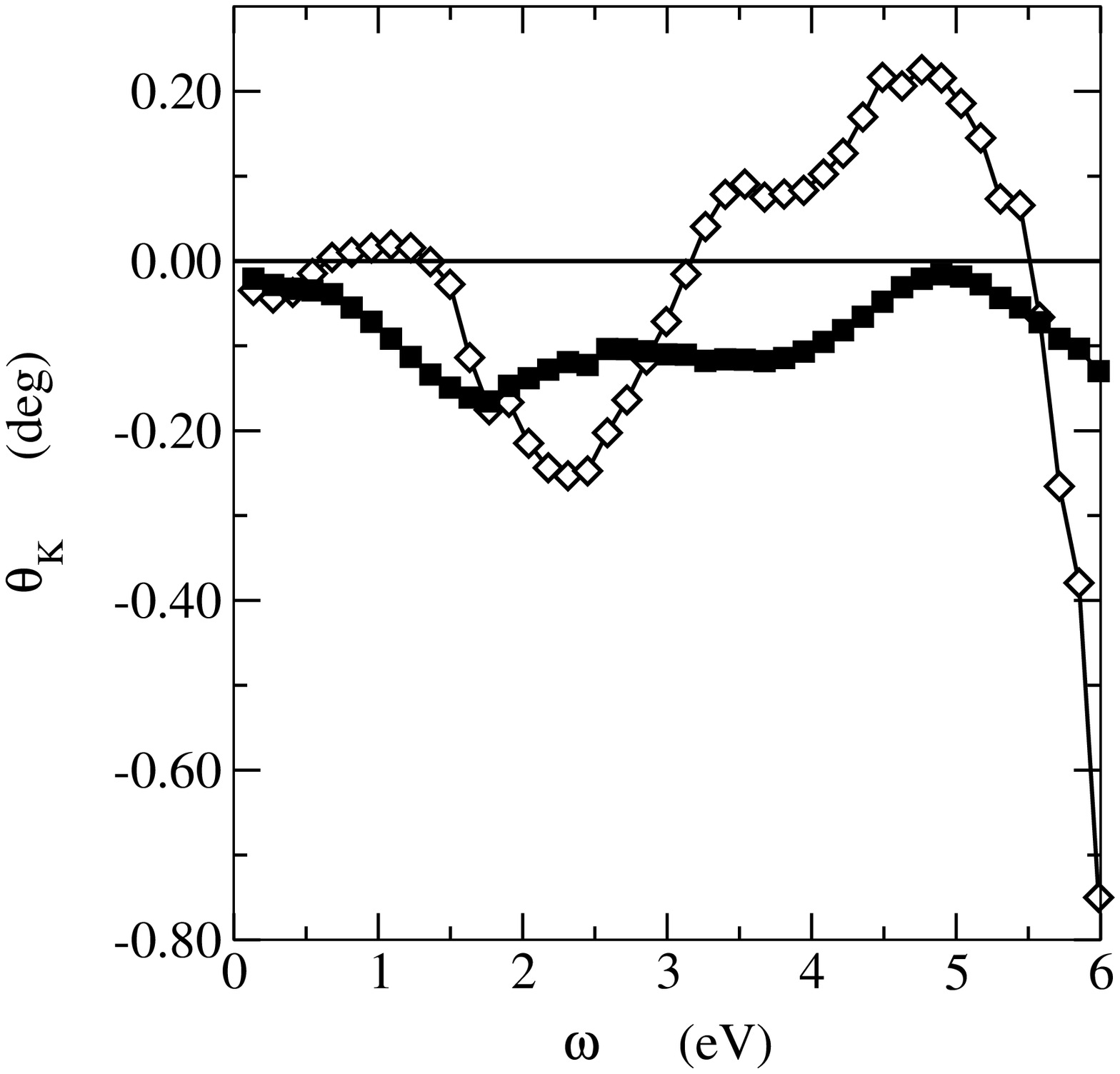} &
\includegraphics[width=0.47\columnwidth,clip]{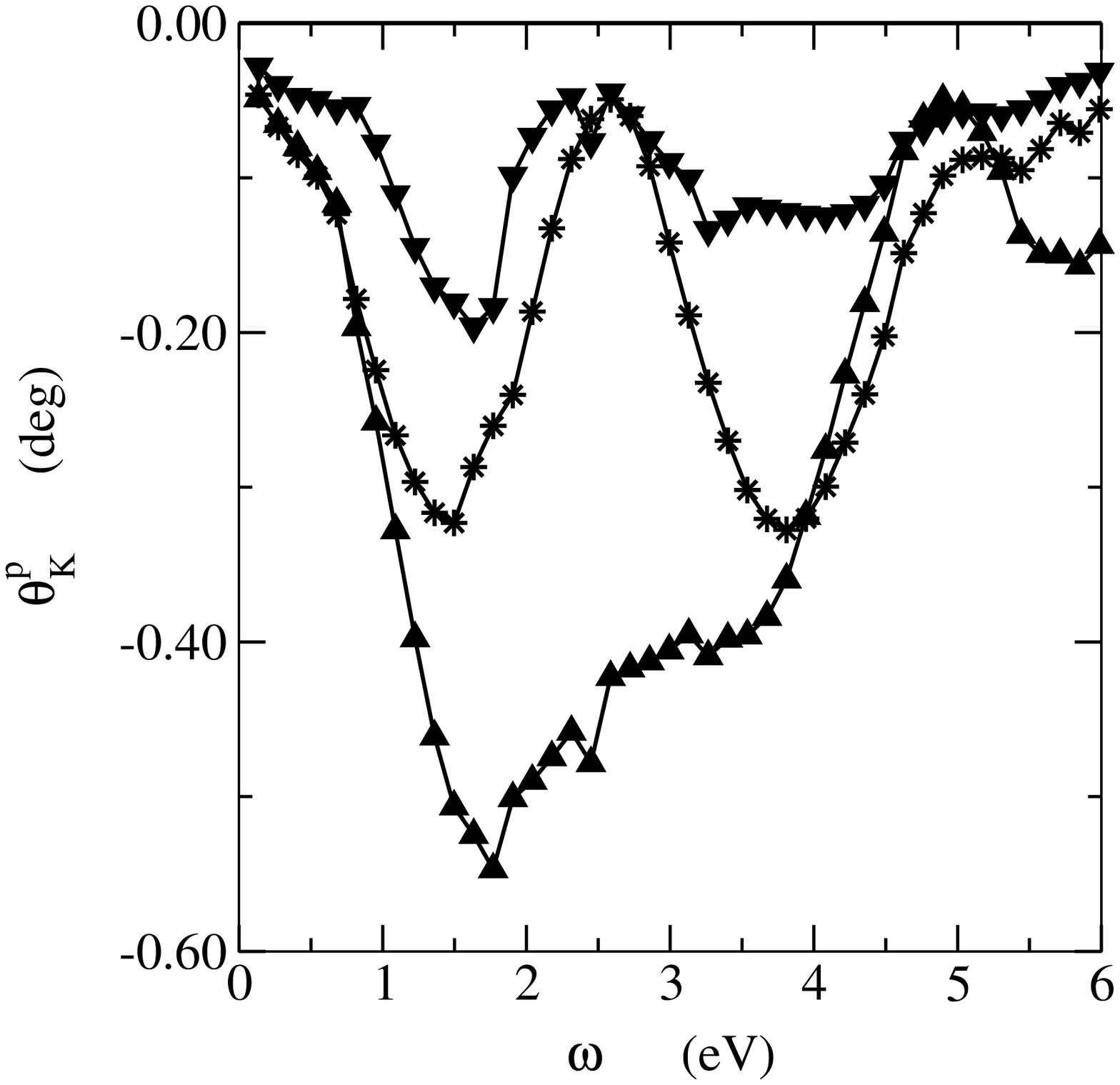} \\
\includegraphics[width=0.47\columnwidth,clip]{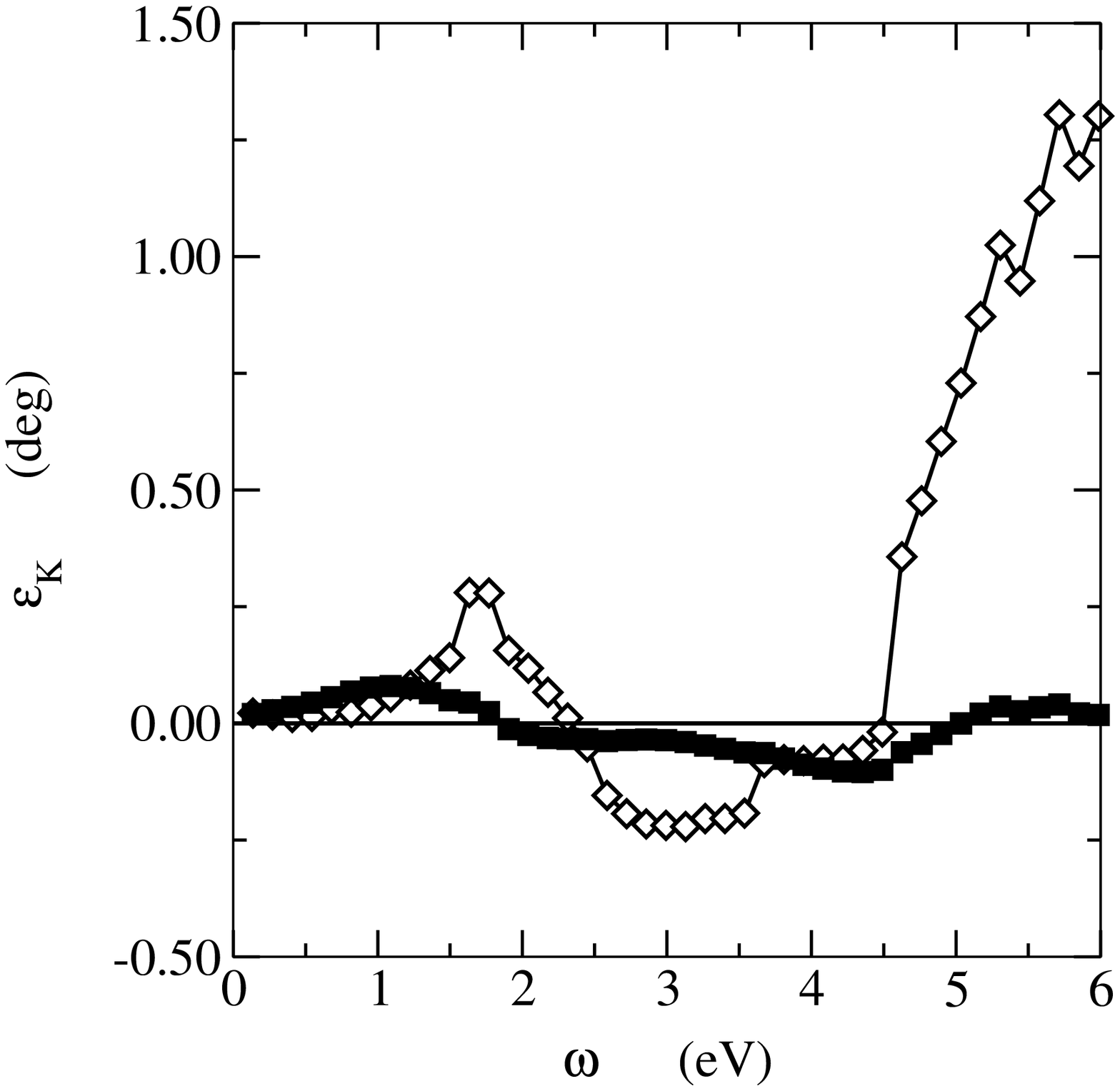} &
\includegraphics[width=0.47\columnwidth,clip]{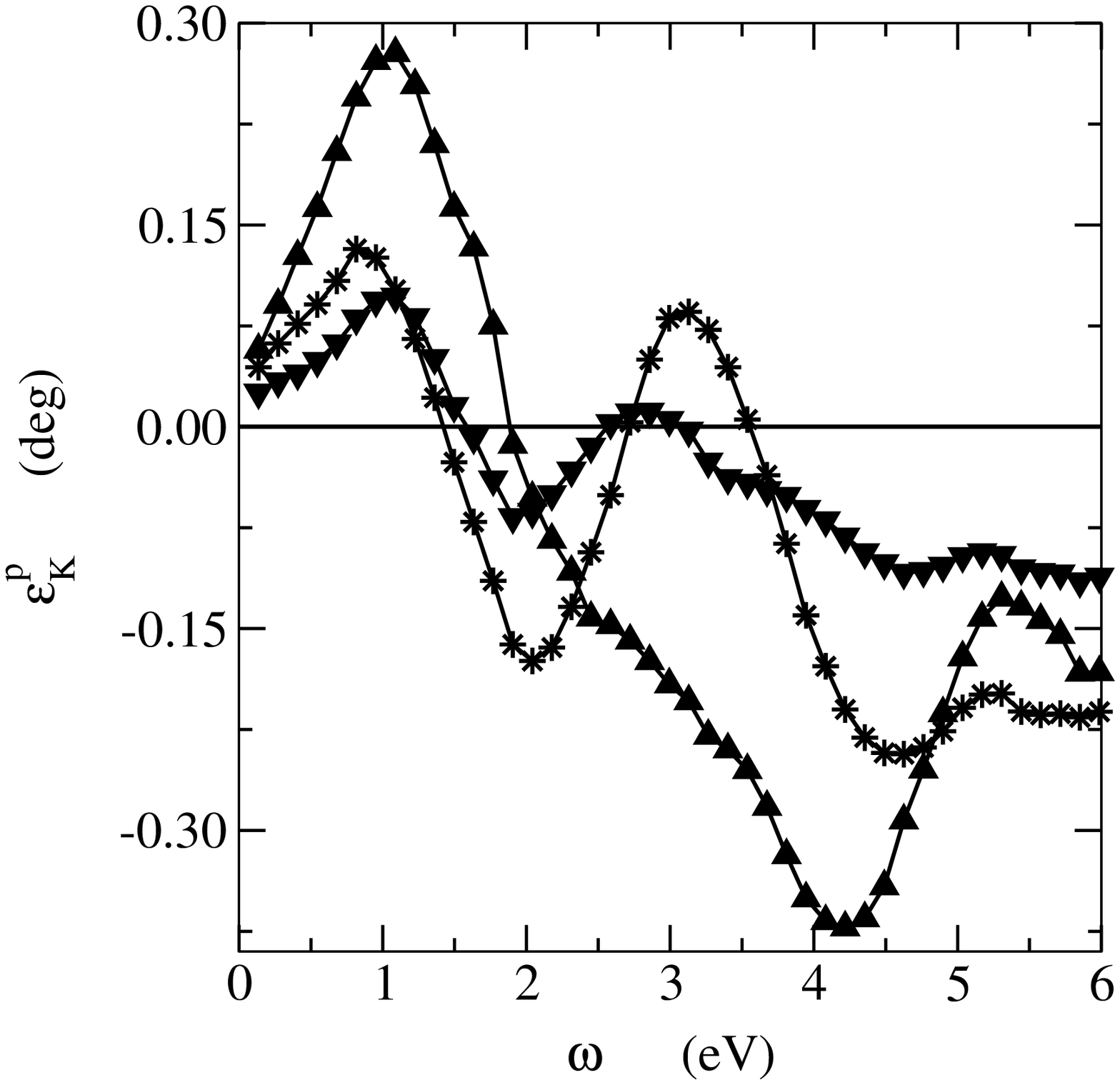} \\
\end{tabular}
\caption[Layer--resolved complex Kerr effect.]
    {\label{fig:lreskerr}
      Layer--resolved complex Kerr effect for
      $\mathrm{Co}\mid\mathrm{Pt}_{m=3}$. In the left panel the Kerr
      spectra resolved for the surface Co--layer
      ($\scriptstyle\Diamond$) are compared with that of the layered
      system ($\scriptstyle\blacksquare$). The right panel show the
      complex Kerr effect arising from the Pt--layers:
      $\blacktriangle$ refers to the Kerr spectra resolved for the
      first Pt--layer below the surface, $\ast$ for the second
      Pt--layer and $\blacktriangledown$ for the last Pt--layer,
      respectively. }
\end{figure}

To understand this, we introduce layer--resolved complex Kerr angles
by using instead of $\Sigma _{\mu\nu} (\omega)$ in Eq.\
(\ref{eq:moke}), the layer--resolved $\Sigma ^{p} _{\mu\nu} (\omega)$,
i.e.,\
\begin{equation}
\Phi ^{p}_{\mathrm{K}} = \theta ^{\,p}_{\mathrm{K}} + i \varepsilon
^{\,p}_{\mathrm{K}} = \dfrac{\Sigma ^{\,p}_{\mathrm{xy}}(\omega)} 
{\Sigma ^{\,p}_{\mathrm{xx}}(\omega)} \left[ 1 + 
\dfrac{4\pi i}{\omega}\,\Sigma ^{\,p}_{\mathrm{xx}}(\omega) \right]
^{-\frac{1}{2}} \quad (p=1,\ldots,n) \ .
\label{eq:lrmoke}
\end{equation}
For $\mathrm{Co}\mid\mathrm{Pt}_{m=3}$, $\Phi ^{p}_{\mathrm{K}}$ is
plotted in Fig.\ \ref{fig:lreskerr}. In contrast to a homogeneous
system, like bcc--Fe or fcc--Co \cite{HE99}, it is found that the
surface resolved MOKE cannot predict correctly the complex Kerr effect
in our inhomogeneous, surface layered system $\mathrm{Co} \mid
\mathrm{Pt} _{m=3}$, see left panels of Fig.\ \ref{fig:lreskerr}.
Inspecting the layer--resolved complex Kerr angles of the three
Pt--layers (right panels in Fig.\ \ref{fig:lreskerr}), it can be seen
that both $\theta ^{\,p}_{\mathrm{K}}$ and $\varepsilon
^{\,p}_{\mathrm{K}}$ of the first Pt--layer below the surface
Co--layer is in the experimental range \cite{Wel96}. This shows that
the first buffer layer is as important as the magnetic surface in an
inhomogeneous layered system. Thus in order to exploit the relatively
big MOKE in the first buffer layer, one has to bring Pt into the
surface Co--layer, e.g.\ by alloying, and has to prepare thin
films. This theoretical finding is completely in--line with the known
experimental facts \cite{HTB97}.

\begin{figure}[htbp] \centering
\begin{tabular}{cc}
\includegraphics[width=0.47\columnwidth,clip]{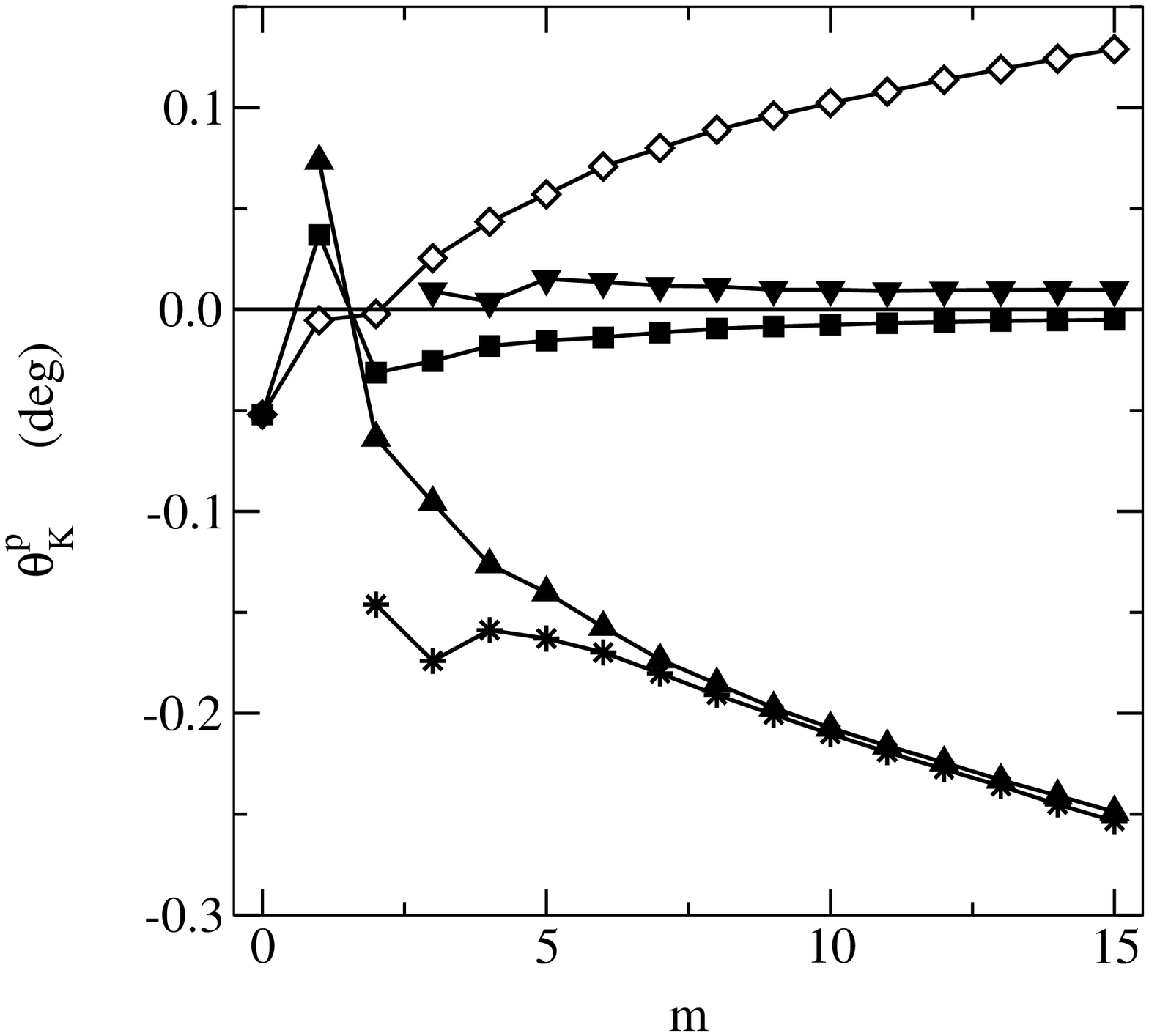} &
\includegraphics[width=0.47\columnwidth,clip]{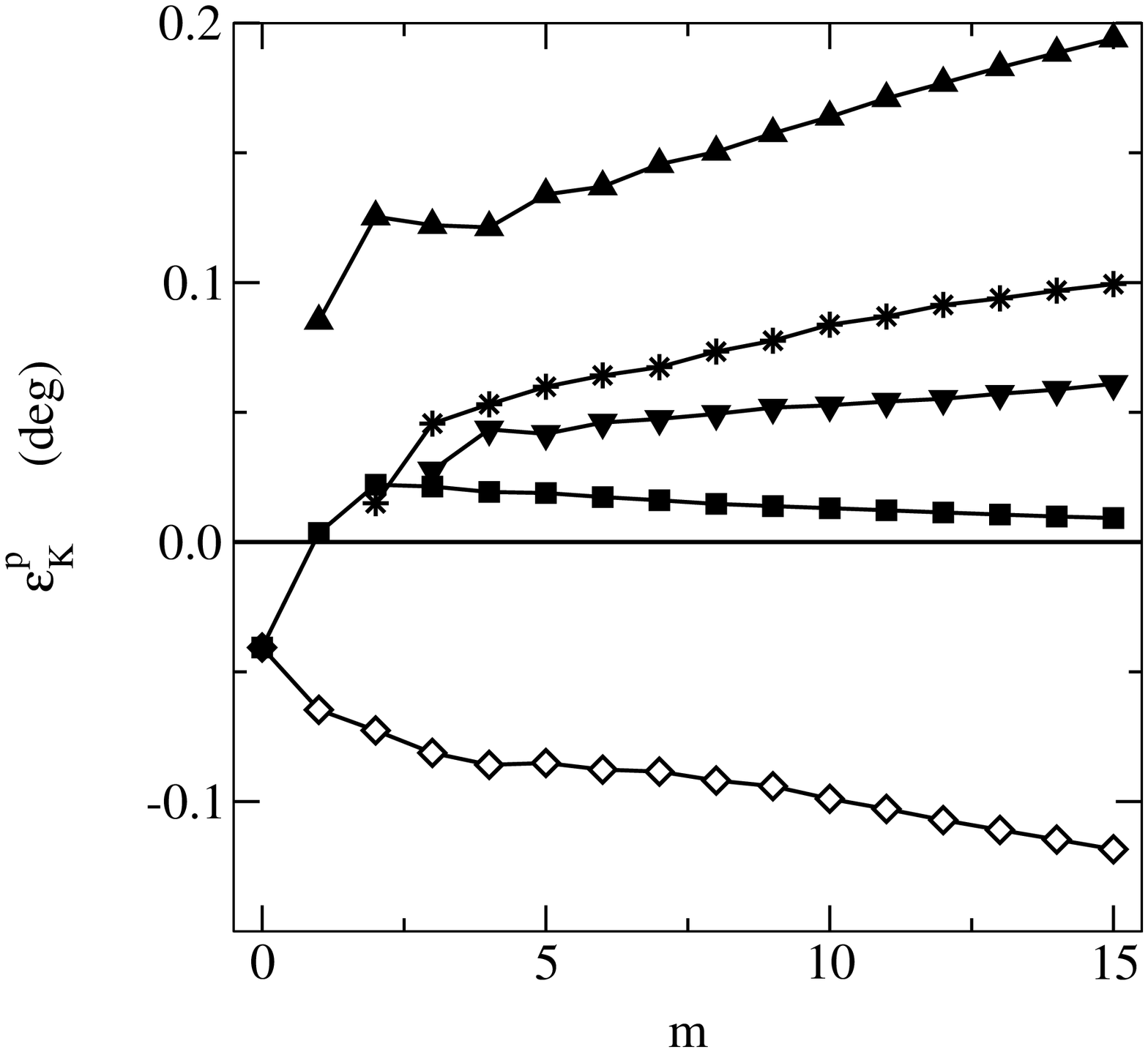} \\
\end{tabular}
\caption[$\protect n$--convergence of the complex Kerr effect
    for $\protect\hbar\omega$ = 0.05 Ryd and $\protect T$ = 300 K.]
    {\label{fig:nconv4kerr}
      Total and layer--resolved complex Kerr angle for $\hbar\omega =
      0.68$ eV and the $\mathrm{Co}\mid\mathrm{Pt}_m$ multilayer
      ($m=0,\ldots,15$), containing below the surface Co--layer six
      self--consistently included Pt--layers followed by bulk--like
      layers. (The used symbols represent the same type of data as in Fig.\
      \ref{fig:lreskerr}.) }
\end{figure}

In addition, we have also studied the variation of MOKE in
$\mathrm{Co} \mid \mathrm{Pt} _m$ with the thickness of the Pt buffer
for $\hbar\omega = 0.68$ eV.  The change in the layer--resolved MOKE
for $m$ up to $15$ is given in Fig.\ \ref{fig:nconv4kerr}. The
$\mathrm{Co} \mid \mathrm{Pt} _m$ system has only six Pt--layers below
the surface included self--consistently and the other Pt--layers are
all bulk--like. This construction is based on the fact, that no
changes occur neither in MOKE nor in the optical conductivity, if the
system has more than six self--consistently included Pt--layers. (Test
calculations were carried out for $\mathrm{Co} \mid \mathrm{Pt}
_{m=9}$ and $\hbar\omega = 0.68$ eV by changing the number of
bulk--like Pt-layers in the system.)

As can be seen from Fig.\ \ref{fig:nconv4kerr}, the layer--resolved
complex Kerr angles up to nine (ten) Pt mono--layers do not depend
linearly on the thickness of the Pt buffer. This is in agreement with
other, ab--initio \cite{BSW+00} and model \cite{BII+99}
magneto--transport calculations ($\omega=0$) for different multilayer
systems, but does not confirm the situation found performing
super--cell calculations \cite{PE00}, where linearity of the surface
resolved optical conductivity seems to occur after a few layers.

We have found, that the Kerr angle of the surface Co--layer increases
with the thickness of Pt, whereas that of the first Pt--layer below
the surface, decreases and the opposite holds for the layer--resolved
Kerr ellipticity. Although, these two layers provide the main
contributions to the MOKE in $\mathrm{Co} \mid \mathrm{Pt}_m$, they
alone are not sufficient to describe the complex Kerr angle of the
whole system (over nine Pt--layers), because at least twelve
Pt--layers are needed to get $\theta_{\mathrm{K}}$ and
$\varepsilon_{\mathrm{K}}$ stabilized around -0.005 and 0.009 deg,
respectively. The change in the surface resolved MOKE with respect to
$m$ is compensated mainly by the change in the layer--resolved MOKE
arising from the first non--magnetic layer below the surface, and
hence $\theta_{\mathrm{K}}$ and $\varepsilon_{\mathrm{K}}$ finally are
converging, for simplicity the layer--resolved MOKE of the Pt--layers
below the third one are not shown in Fig.\ \ref{fig:nconv4kerr}. They
are situated always in--between the results obtained for the surface
Co-- and the first Pt--layer below. The deeper the Pt--layer, the
smaller its MOKE is, such that the contribution arising from a
Pt--layer below the twelve--th layer has a really neglectable
influence on the total MOKE.

\begin{figure}[hbtp] \centering
\begin{tabular}{cc}
\includegraphics[width=0.47\columnwidth,clip]{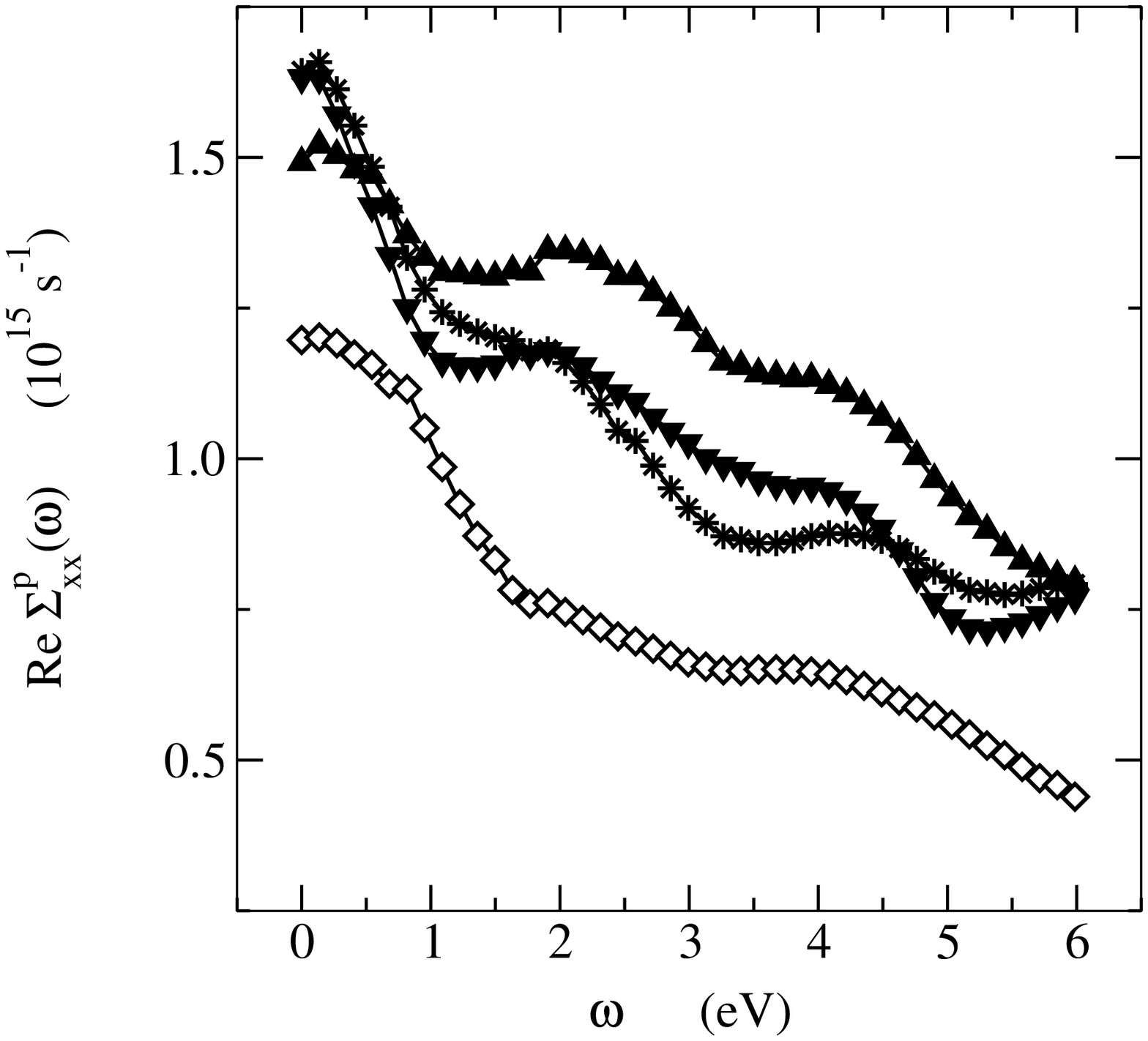} &
\includegraphics[width=0.47\columnwidth,clip]{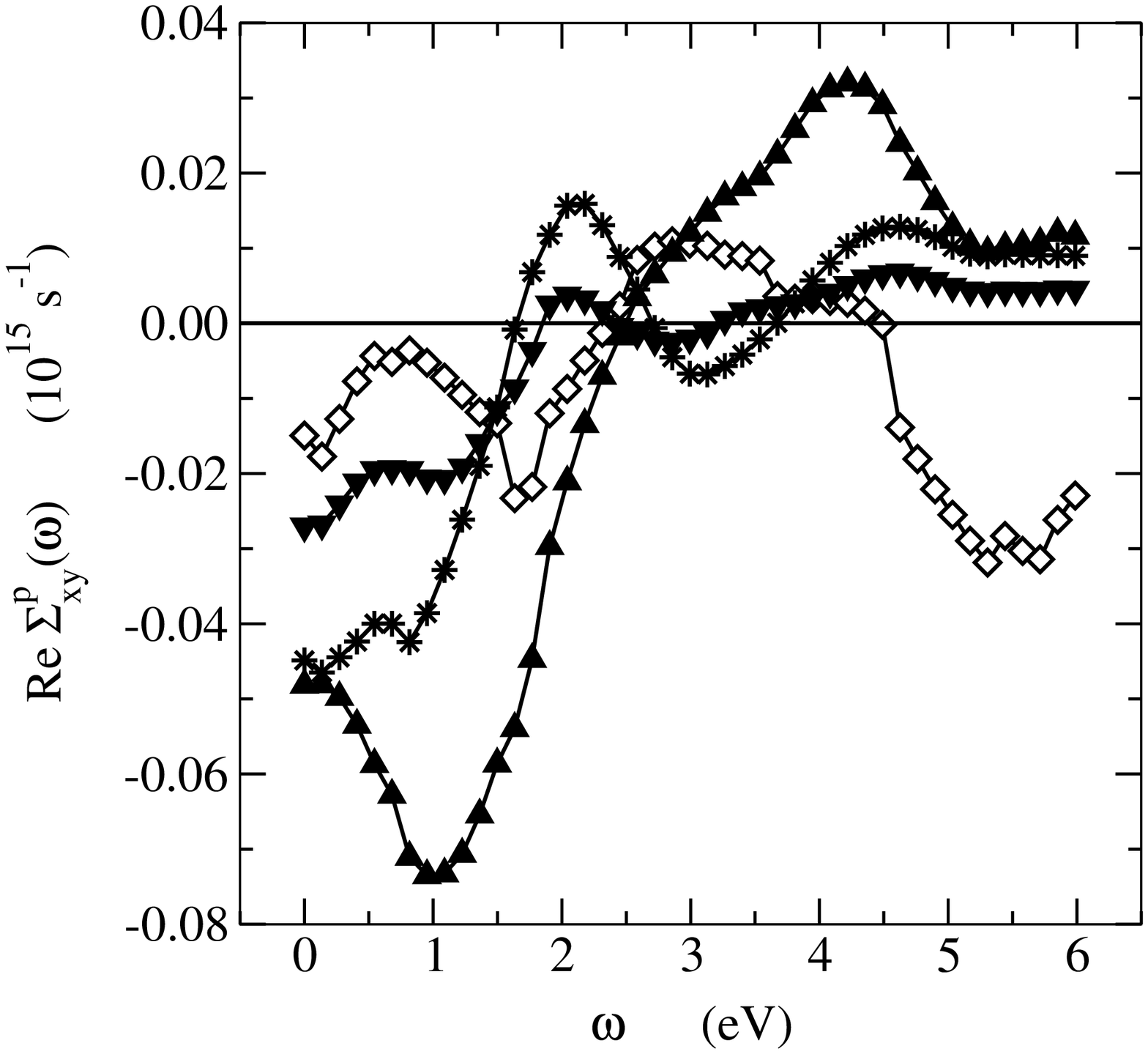} \\
\includegraphics[width=0.47\columnwidth,clip]{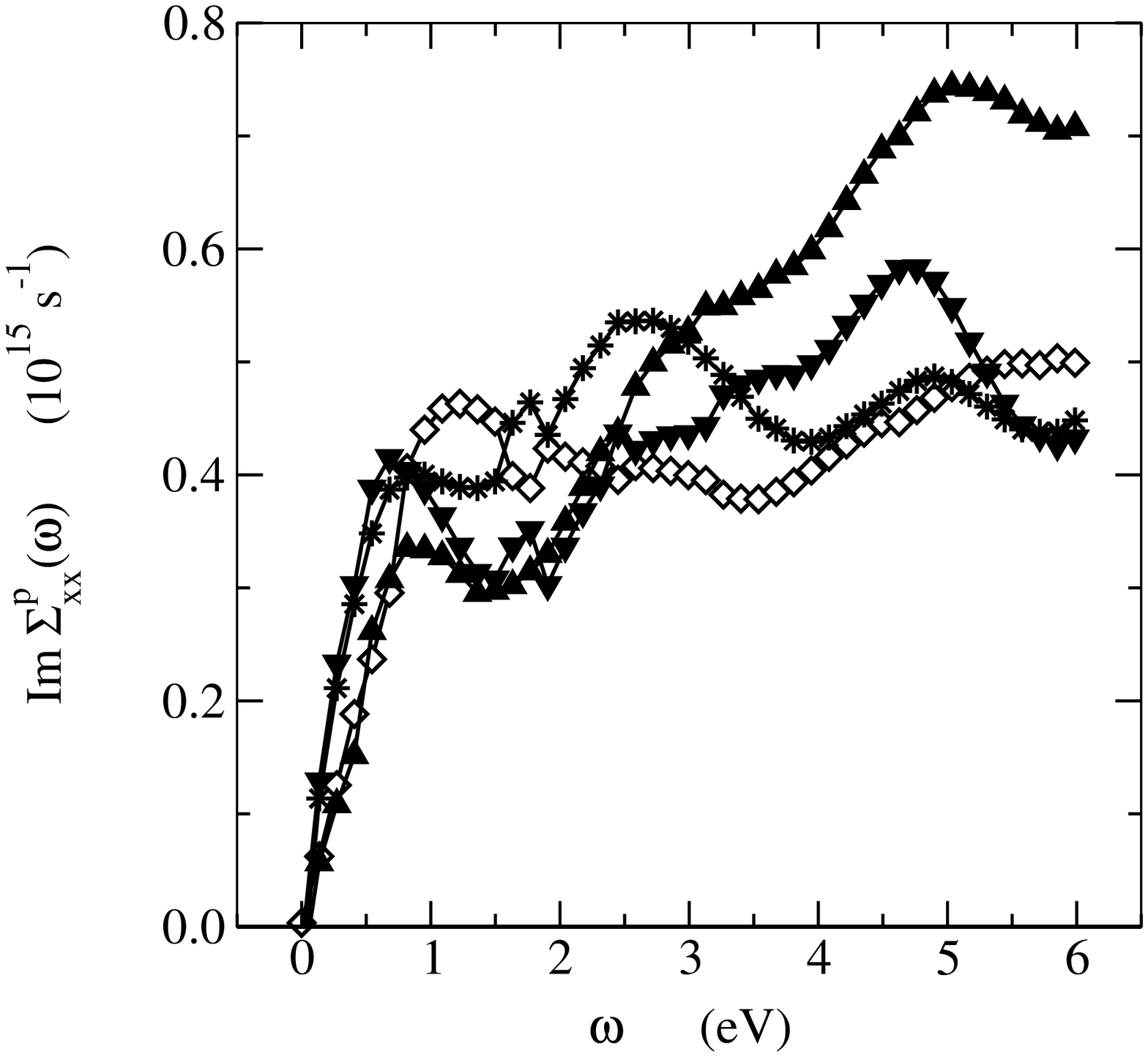} &
\includegraphics[width=0.47\columnwidth,clip]{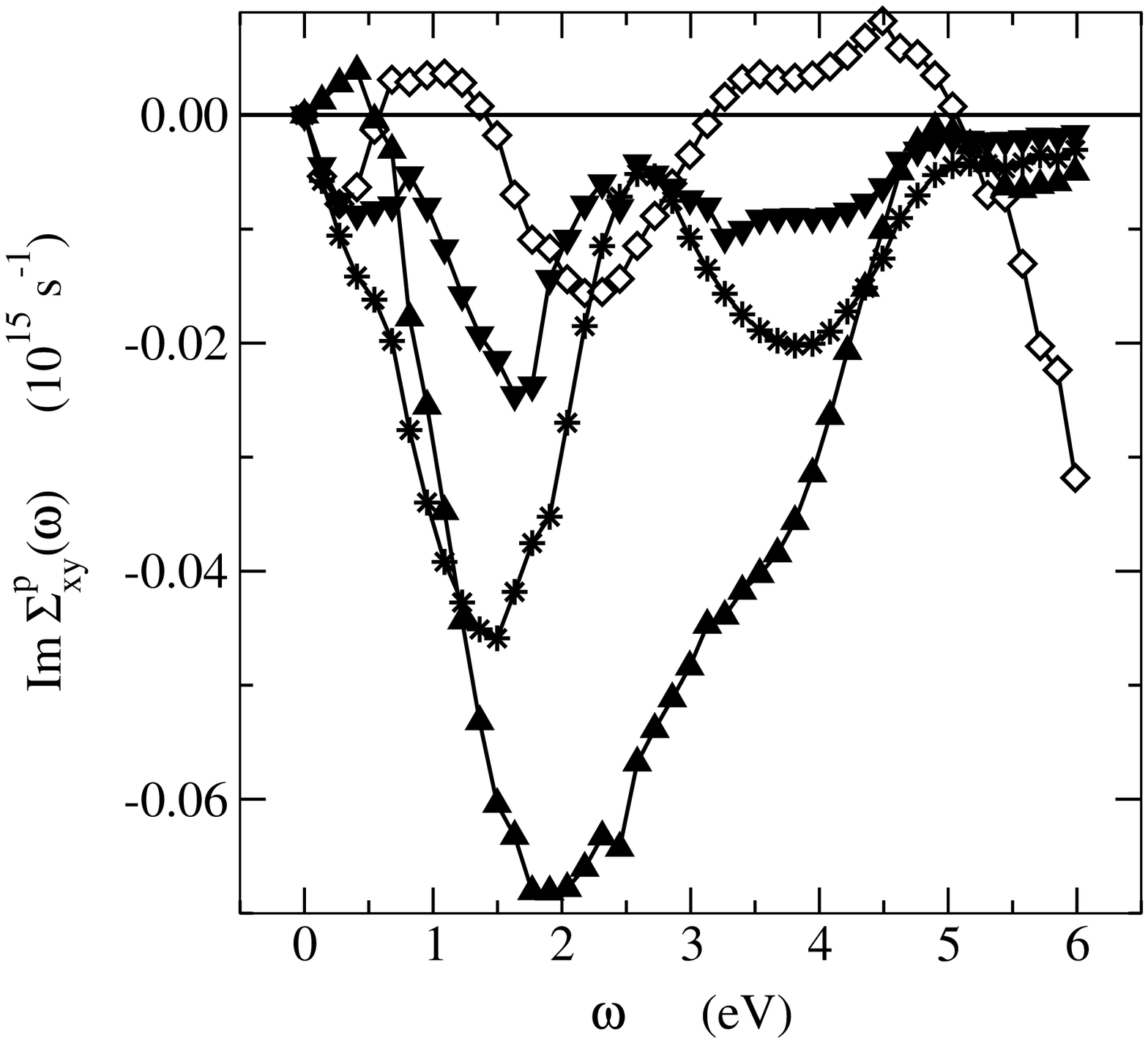}
\end{tabular}
\caption[Layer--resolved optical conductivity.]
    {\label{fig:lressgm}
    Real and imaginary part of the total and layer--resolved optical
    conductivity for $\mathrm{Co}\mid\mathrm{Pt}_{m=3}$.  The optical
    conductivity resolved for the surface Co--layer is given by
    $\scriptstyle\Diamond$ and that resolved for the Pt--layers by
    $\blacktriangle$, $\ast$ and $\blacktriangledown$, respectively.  
    The optical conductivity of the multilayer is represented by $\ast$.}
\end{figure}

Finally, in Fig.\ \ref{fig:lressgm} we show that both, the
layer--resolved and the total optical conductivity tensor,
respectively, can indeed be computed for $\omega=0$ with a finite
life--time broadening by means of the contour integration technique,
see also Section\ \ref{sect:limits}. As can be seen from Fig.\
\ref{fig:lressgm}, the imaginary parts of all $\Sigma ^{p} _{\mu\nu}
(\omega=0)$ and $\Sigma _{\mu\nu} (\omega=0)$ vanish (with an accuracy
of $10^{-3}\, \mathrm{fs}^{-1}$), whereas the real parts of the same
quantities remain finite. However, these values cannot be used to
calculate the MOKE based on Eqs.\ (\ref{eq:moke}) and
(\ref{eq:lrmoke}), because these expressions are diverging in the
static limit.

\section{Summary}
\label{sect:summary}
%

The contour integration technique applied to Luttinger's formalism for
the optical conductivity tensor has been extended to electric fields
with arbitrary polarizations and is shown to straightforwardly account
for both the inter-- and intra--band contribution. Hence within our
technique there is no need to approximate the intra--band contribution
by the so--called semi--empirical Drude term. The optical conductivity
tensor for finite life--time broadening can be computed also in the
zero frequency limit, however, the dc conductivity cannot be obtained
integrating in the complex energy plane. The contour integration
technique was also completed here in deriving formulae for the
current--current correlation function at zero frequency and vanishing
life--time broadening ($\zeta=0$).

Introducing the concept of the layer--resolved complex Kerr angles and
computing the magneto--optical Kerr effect for the $\mathrm{Co} \mid
\mathrm{Pt} _m$ surface layered system in polar geometry, we have
shown that (1) the layer--resolved complex Kerr angle for the
ferromagnetic surface alone cannot predict correctly the
magneto--optical Kerr effect in inhomogeneous multilayers. In the case
of $\mathrm{Co} \mid \mathrm{Pt} _m$ the first Pt--layer below the
surface is as important for the polar MOKE as the ferromagnetic
surface Co--layer itself.  (2) The Kerr spectra of $\mathrm{Co} \mid
\mathrm{Pt} _{m=3}$ possess all the known, experimental
characteristics, but in order to exploit the relatively big complex
Kerr angle arising from the first Pt--layer below the surface, one has
to consider a Co--Pt alloy at the surface.

In addition, it was also shown that the layer--resolved MOKE shows a
non--linear dependence with respect to the thickness of the buffer Pt
up to $m=9$ for $\mathrm{Co} \mid \mathrm{Pt} _m$; only for $m\ge 12$
the complex Kerr angle is fully converged.

Finally, we have demonstrated that the contour integration technique
developed for the Luttinger formalism can be used straightforwardly
with a finite life--time broadening in the zero frequency limit.
Calculations performed provide a purely real optical conductivity
tensor at zero frequency.

%
%
\section{Acknowledgements}
%
This work was supported by the Austrian Ministry of Science (Contract
No. 45.451/1-III/B/8a/99) and by the Research and Technological
Cooperation Project between Austria and Hungary (Contract
No. A-35/98).  One of the authors (L.S.) is also indebted to partial
support by the Hungarian National Science Foundation (Contract No.
OTKA T030240).
%
%
%

%
%
\newpage
%
 
%
\newpage 
%
 
%
\newpage 
%
 
%
\newpage
%
 
%
\newpage
%
 
%

\begin{thebibliography}{45}

\expandafter\ifx\csname natexlab\endcsname\relax\def\natexlab#1{#1}\fi
\expandafter\ifx\csname url\endcsname\relax
\def\url#1{{\tt #1}}\fi
\expandafter\ifx\csname urlprefix\endcsname\relax\def\urlprefix{URL }\fi

\bibitem[{Abramowitz and Stegun(1972)}]{AS72}
Abramowitz, M. and I.A.\ Stegun (1972).
\newblock {\em Handbook of Mathematical Functions with Formulas, Graphs and
  Mathematical Tables}.
\newblock Dover, New York.

\bibitem[{Bruno {\em et al.\/}(1999)Bruno, Itoh, Inoue and Nonoya}]{BII+99}
Bruno, P., H.\ Itoh, J.\ Inoue and S.\ Nonoya (1999).
\newblock Influence of disorder on the perpendicular magnetoresistance of
  magnetic multilayers.
\newblock {\em J. Magn. Magn. Materials}, {\bf 198--199}, 46.

\bibitem[{Bertero and Sinclair(1994)}]{BS94}
Bertero, G.A. and R.\ Sinclair (1994).
\newblock Structure--property correlations in Pt/Co multilayers for
  magneto--optic recording.
\newblock {\em J. Magn. Magn. Materials}, {\bf 134}, 173.

\bibitem[{Blaas {\em et al.\/}(2000)Blaas, Szunyogh, Weinberger,
Sommers and Levy}]{BSW+00}
Blaas, C., L.\ Szunyogh, P.\ Weinberger, C.\ Sommers and P.M.\ Levy
(2000). 
\newblock Electrical transport properties of bulk Ni$_c$Fe$_{1-c}$ alloys and
  related spin--valve systems.
\newblock {\em Phys. Rev. B}, manuscript no. BU7531.

\bibitem[{Butler(1985)}]{But85}
Butler, W.H. (1985).
\newblock Theory of electronic transport in random alloys:
  Korringa--Kohn--Rostoker coherent--potential approximation.
\newblock {\em Phys. Rev. B}, {\bf 31}, 3260.

\bibitem[{Blaas {\em et al.\/}(1999)Blaas, Weinberger, Szunyogh, Levy
and Sommers}]{CWS+99}
Blaas, C., P.\ Weinberger, L.\ Szunyogh, P.M.\ Levy and C.B.\ Sommers (1999).
\newblock Ab initio calculations of magnetotransport for magnetic multilayers.
\newblock {\em Phys. Rev. B}, {\bf 60}, 492.

\bibitem[{Butler {\em et al.\/}(1994)Butler, Zhang, Nicholson and
MacLaren}]{BZN+94}
Butler, W.H., X.G.\ Zhang, D.M.C.\ Nicholson and J.M.\ MacLaren (1994).
\newblock Theory of transport in inhomogeneous systems and application to
  magnetic multilayer systems.
\newblock {\em J. Appl. Physics}, {\bf 76}, 6808.

\bibitem[{Callaway(1974)}]{Cal74}
Callaway, J. (1974).
\newblock {\em Quantum Theory of the Solid State part B}.
\newblock Academic Press, New York.

\bibitem[{Calvetti {\em et al.\/}(2000)Calvetti, Golub, Gragg and
Reichel}]{CGG+00}
Calvetti, D., G.H.\ Golub, W.B.\ Gragg and L.\ Reichel (2000).
\newblock Computation of Gauss--Konrod quadrature rules.
\newblock {\em Math. Comput.}, {\bf 69}, 1035.

\bibitem[{Daalderop {\em et al.\/}(1988)Daalderop, Mueller, Albers and
Boring}]{DMAB88}
Daalderop, G.H.O., F.M.\ Mueller, R.C.\ Albers and A.M.\ Boring (1988).
\newblock Prediction of a large polar Kerr angle in NiUSn.
\newblock {\em Appl. Physics Lett.}, {\bf 52}, 1636.

\bibitem[{Ebert {\em et al.\/}(1998)Ebert, Perlov, Yaresko, Antonov
and Uba}]{EPY+98}
Ebert, H., A.\ Perlov, A.N.\ Yaresko, V.N.\ Antonov and S.\ Uba (1998).
\newblock Theoretical investigations on the magneto-optical properties of
  transition metal multilayer and surface systems.
\newblock {\em Mat. Res. Soc. Symp. Proc.}, {\bf 475}, 407.

\bibitem[{Eykholt(1986)}]{Eyk86}
Eykholt, R. (1986).
\newblock Extension of the Kubo formula for the electrical conductivity tensor
  to arbitrary polarizations of the electric field.
\newblock {\em Phys. Rev. B}, {\bf 34}, 6669.

\bibitem[{Gao {\em et al.\/}(1998)Gao, DeVries, Thompson and Woollam}]{GVT+98}
Gao, X., M.J.\ DeVries, D.W.\ Thompson and J.A.\ Woollam (1998).
\newblock Thickness dependence of interfacial magneto--optic effects in
  Pt$\mid$Co multilayers.
\newblock {\em J. Appl. Physics}, {\bf 83}, 6747.

\bibitem[{Guo and Ebert(1995)}]{GE95}
Guo, G.Y. and H.\ Ebert (1995).
\newblock Band theoretical investigation of the magneto-optical Kerr effect in
  Fe and Co multilayers.
\newblock {\em Phys. Rev. B}, {\bf 51}, 12633.

\bibitem[{Greenwood(1958)}]{Gre58}
Greenwood, D.A. (1958). 
\newblock The Boltzmann equation in the theory of electrical conduction in
  metals.
\newblock {\em Proc. Phys. Soc.}, {\bf 71}, 585.

\bibitem[{Huhne and Ebert(1999)}]{HE99}
Huhne, T. and H. Ebert (1999).
\newblock Fully relativistic description of the magneto--optical properties of
  arbitrary layered systems.
\newblock {\em Phys. Rev. B}, {\bf 60}, 12982.

\bibitem[{Hatwar {\em et al.\/}(1997)Hatwar, Tyan and Brucker}]{HTB97}
Hatwar, T.K., Y.S.\ Tyan and C.F.\ Brucker (1997).
\newblock High--performance Co/Pt multilayer magneto--optical disk using
  ultrathin seed layers.
\newblock {\em J. Appl. Physics}, {\bf 81}, 3839.

\bibitem[{Yu-Kuang Hu(1993)}]{Hu93}
Yu-Kuang Hu, B. (1993).
\newblock Simple derivation of a general relationship between imaginary- and
  real-time Green's correlation functions.
\newblock {\em Am. J. Phys.}, {\bf 61}, 457.

\bibitem[{Hama and Watanabe(1992)}]{Haw92}
Hama, J. and M.\ Watanabe (1992).
\newblock General formulae for the special points and their weighting in {\bf
  k}-space integration.
\newblock {\em J. Phys.: Condensed Matter}, {\bf 4}, 4583.

\bibitem[{Kubo(1957)}]{Kub57}
Kubo, R. (1957).
\newblock Statistical--mechanical theory of irreversible processes I.
\newblock {\em J. Phys. Soc. Japan}, {\bf 12}, 570.

\bibitem[{Kubo(1966)}]{Kub66}
Kubo, R. (1966).
\newblock The fluctuation--dissipation theorem.
\newblock {\em Rep. Prog. Phys.}, {\bf 29}, 255.

\bibitem[{Laurie(1997)}]{Lau97}
Laurie, D.P. (1997).
\newblock Calculation of Gauss--Konrod quadrature rules.
\newblock {\em Math. Comput.}, {\bf 66}, 1133.

\bibitem[{Lax(1958)}]{Lax58}
Lax, M. (1958).
\newblock Generalized mobility theory.
\newblock {\em Phys. Rev.}, {\bf 109}, 1921.

\bibitem[{Luttinger(1964)}]{Lut64}
Luttinger, J.M. (1964).
\newblock Theory of thermal transport coefficients.
\newblock {\em Phys. Rev.}, {\bf 135A}, 1505.

\bibitem[{Luttinger(1967)}]{Lut67}
Luttinger, J.M. (1967).
\newblock Transport theory.
\newblock In {\em Mathematical Methods in Solid State and Superfluid Theory},
\newblock Oliver and Boyd, Edingburgh, Chap. 4, pp. 157.

\bibitem[{Mahan(1990)}]{Mah90}
Mahan, G.D. (1990).
\newblock {\em Many--Particle Physics}.
\newblock Plenum Press, New York.

\bibitem[{Mansuripur(1995)}]{Man95}
Mansuripur, M. (1995).
\newblock {\em The Principles of Magneto-Optical Recording}.
\newblock Cambridge University Press, Cambridge.

\bibitem[{Nicholson {\em et al.\/}(1994)Nicholson, Stocks, Wang,
Shelton, Szotek and Temmerman}]{NSW+94}
Nicholson, D.M.C., G.M.\ Stocks, Y.\ Wang, W.A.\ Shelton, Z.\ Szotek
and W.M.\ Temmerman (1994).
\newblock Stationary nature of the density--functional free energy: Application
  to accelerated multiple--scattering calculations.
\newblock {\em Phys. Rev. B}, {\bf 50}, 14686.

\bibitem[{Oppeneer {\em et al.\/}(1992)Oppeneer, Maurer, Sticht and
K\"{u}bler}]{OMSK92}
Oppeneer, P.M., T.\ Maurer, J.\ Sticht and J.\ K\"ubler (1992).
\newblock Ab initio calculated magneto--optical Kerr effect of ferromagnetic
  metals: Fe and Ni.
\newblock {\em Phys. Rev. B}, {\bf 45}, 10924.

\bibitem[{Oppeneer(1999)}]{Opp99}
Oppeneer, P.M. (1999).
\newblock Theory of the magneto--optical Kerr effect in ferromagnetic
  compounds.
\newblock Habilitationsschrift, Technische Universit\"{a}t Dresden.

\bibitem[{Perlov and Ebert(2000)}]{PE00}
Perlov, A.Ya. and H.\ Ebert (2000).
\newblock Layer--resolved optical conductivity of magnetic multilayers.
\newblock {\em Europhys. Lett.}, {\bf 52}, 108.

\bibitem[{Press {\em et al.\/}(1992)Press, Flannery, Teukolsky and
Vetterling}]{PFT+92}
Press, W.H., B.P.\ Flannery, S.A.\ Teukolsky and W.T.\ Vetterling (1992).
\newblock {\em Numerical Recipes in Fortran: The Art of Scientific Computing}.
\newblock Cambridge University Press, Cambridge.

\bibitem[{Pustogowa {\em et al.\/}(1999)Pustogowa, Zabloudil,
Uiberacker, Blaas, Weinberger, Szunyogh and Sommers}]{PZU+99}
Pustogowa, U., J.\ Zabloudil, C.\ Uiberacker, C.\ Blaas, P.\
Weinberger, L.\ Szunyogh and C.\ Sommers (1999).
\newblock Magnetic properties of thin films of Co and of (CoPt) superstructures
  on Pt(100) and Pt(111).
\newblock {\em Phys. Rev. B}, {\bf 60}, 414.

\bibitem[{Reim and Schoenes(1990)}]{RS90}
Reim, W. and J.\ Schoenes (1990).
\newblock Magneto--optical spectroscopy of $f$--electron systems.
\newblock In K.H.J.\ Buschow and E.P.\ Wohlfarth (Eds.), {\em
  Ferromagnetic Materials},  
\newblock Vol. 5, North-Holland, Amsterdam, Chap. 2, pp. 133.

\bibitem[{Szunyogh {\em et al.\/}(1995)Szunyogh, \'{U}jfalussy and
Weinberger}]{SUW95}
Szunyogh, L., B.\ \'{U}jfalussy and P.\ Weinberger (1995).
\newblock Magnetic anisotropy of iron multilayers on Au(001).
\newblock {\em Phys. Rev. B}, {\bf 51}, 9552.

\bibitem[{Szunyogh {\em et al.\/}(1994)Szunyogh, \'{U}jfalussy
Weinberger and Koll\'{a}r}]{SUWK94}
Szunyogh, L., B.\ \'{U}jfalussy, P.\ Weinberger and J.\ Koll\'{a}r (1994).
\newblock Self--consistent localized KKR scheme for surfaces and interfaces.
\newblock {\em Phys. Rev. B}, {\bf 49}, 2721.

\bibitem[{Szunyogh and Weinberger(1999)}]{SW99}
Szunyogh, L. and P.\ Weinberger (1999).
\newblock Evaluation of the optical conductivity tensor in terms of contour
  integrations.
\newblock {\em J. Phys.: Condensed Matter}, {\bf 11}, 10451.

\bibitem[{\'{U}jfalussy {\em et al.\/}(1995)\'{U}jfalussy, Szunyogh and Weinberger}]{USW95}
\'{U}jfalussy, B., L.\ Szunyogh and P.\ Weinberger (1995).
\newblock Magnetism of 4d and 5d adlayers on Ag(001) and Au(001): comparison
  between a non--relativistic and a fully relativistic approach.
\newblock {\em Phys. Rev. B}, {\bf 51}, 12836.

\bibitem[{Uba {\em et al.\/}(1996)Uba, Uba, Yaresko, Perlov, Antonov
and Gontarz}]{UUY+96}
Uba, S., L.\ Uba, A.N.\ Yaresko, A.Ya.\ Perlov, V.N.\ Antonov and R.\
Gontarz (1996).
\newblock Optical and magneto--optical properties of Co/Pt multilayers.
\newblock {\em Phys. Rev. B}, {\bf 53}, 6526.

\bibitem[{Vernes {\em et al.\/}(2000)Vernes, Szunyogh and Weinberger}]{VSW00a}
Vernes, A., L. Szunyogh and P. Weinberger (2000).
\newblock Numerically improved computational scheme for the optical
  conductivity tensor in layered systems.
\newblock submitted to {\em J. Phys.: Condensed Matter}.

\bibitem[{Wang and Callaway(1974)}]{WC74}
Wang, C.S. and J.\ Callaway (1974).
\newblock Band structure of nickel: spin-orbit coupling, the Fermi surface, and
  the optical conductivity.
\newblock {\em Phys. Rev. B}, {\bf 9}, 4897.

\bibitem[{Weinberger(1990)}]{Wei90}
Weinberger, P. (1990).
\newblock {\em Electron Scattering Theory for Ordered and Disordered Matter}.
\newblock Oxford University Press, Oxford.

\bibitem[{Weller(1996)}]{Wel96}
Weller, D. (1996).
\newblock Magneto--optical Kerr spectroscopy of transition metal alloy
  and compound films. 
\newblock In H.\ Ebert and G. Sch\"{u}tz (Eds.) 
  {\em Spin-orbit Influenced Spectroscopies of Magnetic Solids}, Vol.
  466 of {\em Lecture Notes in Physics}, 
\newblock Springer, Berlin, pp. 1.

\bibitem[{Weinberger {\em et al.\/}(1996)Weinberger, Levy, Banhart,
Szunyogh and \'{U}jfalussy}]{WLB+96}
Weinberger, P., P.M.\ Levy, J.\ Banhart, L.\ Szunyogh and B.\
\'{U}jfalussy (1996).
\newblock Band structure and electrical conductivity of disordered layered
  systems.
\newblock {\em J. Phys.: Condensed Matter}, {\bf 8}, 7677.

\bibitem[{Wildberger {\em et al.\/}(1995)Wildberger, Lang, Zeller and Dederichs}]{WLZ+95}
Wildberger, K., P.\ Lang, R.\ Zeller and P.H.\ Dederichs (1995).
\newblock Fermi--Dirac distribution in ab initio Green's--function
  calculations.
\newblock {\em Phys. Rev. B}, {\bf 52}, 11502.

\end{thebibliography}
\end{document}